\documentclass[onecolumn,nofigure]{pasj01}
\Received{$\langle$reception date$\rangle$}
\Accepted{$\langle$acception date$\rangle$}
\Published{$\langle$publication date$\rangle$}
\usepackage{times}
\usepackage[pdftex]{graphicx,color}

\begin{document}

\title{Optical and X-ray observations of stellar flares on an active M dwarf AD Leonis with Seimei Telescope, SCAT, NICER and OISTER}
%H$\alpha$ Line Broadening, its Relation with Optical Continuum, Rotational Modulation
\author{Kosuke Namekata\altaffilmark{1} }
\author{Hiroyuki Maehara\altaffilmark{2} }
\author{Ryo Sasaki\altaffilmark{3, 4} }
\author{Hiroki Kawai\altaffilmark{3} }
\author{Yuta Notsu\altaffilmark{5, 6} }
\author{Adam F. Kowalski\altaffilmark{5, 6} }
\author{Joel C. Allred\altaffilmark{7} }
\author{Wataru Iwakiri\altaffilmark{3, 4} }
\author{Yohko Tsuboi\altaffilmark{3} }
\author{Katsuhiro L. Murata\altaffilmark{8} }
\author{Masafumi Niwano\altaffilmark{8} }
\author{Kazuki Shiraishi\altaffilmark{8} }
\author{Ryo Adachi\altaffilmark{8} }
\author{Kota Iida\altaffilmark{8} }
\author{Motoki Oeda\altaffilmark{8} }
\author{Satoshi Honda\altaffilmark{9}}
\author{Miyako Tozuka\altaffilmark{9}}
\author{Noriyuki Katoh\altaffilmark{9, 10}}
\author{Hiroki Onozato\altaffilmark{9}}
\author{Soshi Okamoto\altaffilmark{1}}
\author{Keisuke Isogai\altaffilmark{11}}
\author{Mariko Kimura\altaffilmark{1, 4}}
\author{Naoto Kojiguchi\altaffilmark{1}}
\author{Yasuyuki Wakamatsu\altaffilmark{1}}
\author{Yusuke Tampo\altaffilmark{1}}
\author{Daisaku Nogami\altaffilmark{1} }
\author{Kazunari Shibata\altaffilmark{1,12}}

%\altaffiltext{1}{Astronomical Society of Japan, c/o National Astoronomical Observatory of Japan, 2-21-1 Osawa, Mitaka, Tokyo 181-8588, Japan }
\altaffiltext{1}{Department of Astronomy, Kyoto University, Kitashirakawa-Oiwake-cho, Sakyo, Kyoto 606-8502, Japan; namekata@kusastro.kyoto-u.ac.jp}

\altaffiltext{2}{Okayama Branch Office, Subaru Telescope, National Astronomical Observatory of Japan, NINS, Kamogata, Asakuchi, Okayama 719-0232, Japan}

\altaffiltext{3}{Department of Physics, Faculty of Science and Engineering, Chuo University, 1-13-27 Kasuga, Bunkyo-ku, Tokyo 112-8551, Japan}
%\altaffiltext{5}{Okayama Observatory, Kyoto University, 3037-5 Honjo, Kamogata, Asakuchi, Okayama 719-0232, Japan}

\altaffiltext{4}{Institute of Physical and Chemical Research (RIKEN), 2-1 Hirosawa, Wako, Saitama 351-0198, Japan}

\altaffiltext{5}{Laboratory for Atmospheric and Space Physics, University of Colorado Boulder, 3665 Discovery Drive, Boulder, CO 80303, USA }

\altaffiltext{6}{National Solar Observatory, 3665 Discovery Drive, Boulder, CO 80303, USA}

\altaffiltext{7}{NASA/Goddard Space Flight Center, Code 671, Greenbelt, MD 20771, USA}

\altaffiltext{8}{Department of Physics, Tokyo Institute of Technology, 2-12-1 Ookayama, Meguro-ku, Tokyo 152-8551, Japan}

\altaffiltext{9}{Nishi-Harima Astronomical Observatory, Center for Astronomy, University of Hyogo, Sayo, Sayo, Hyogo 679-5313, Japan.}

\altaffiltext{10}{Graduate School of Human Development and Environment, Kobe University, 3-11 Tsurukabuto, Nada-ku, Kobe 657-8501, Japan}

\altaffiltext{11}{Okayama Observatory, Kyoto University, 3037-5 Honjo, Kamogatacho, Asakuchi, Okayama 719-0232, Japan}

\altaffiltext{12}{Astronomical Observatory, Kyoto University, Kitashirakawa-Oiwake-cho, Sakyo, Kyoto 606-8502, Japan.}

%\altaffiltext{12}{JSPS Overseas Research Fellow}

\email{namekata@kusastro.kyoto-u.ac.jp}


\maketitle

\begin{abstract}
We report multi-wavelength monitoring observations of an M-dwarf flare star AD Leonis with Seimei Telescope (6150--7930 {\AA}), SCAT (Spectroscopic Chuo-university Astronomical Telescope; 3700--7500 {\AA}), NICER (Neutron Star Interior Composition Explorer; 0.2--12.0 keV), and collaborations of OISTER (Optical and Infrared Synergetic Telescopes for Education and Research) program.
Twelve flares are detected in total which include ten H$\alpha$, four X-ray, and four optical-continuum flares; one of them is a superflare with the total energy of $\sim$ 2.0$\times$10$^{33}$ erg.
We found that (1) during the superflare, the H$\alpha$ emission line  full width at 1/8 maximum  dramatically increases to 14 {\AA} from 8 {\AA} in the low-resolution spectra (R$\sim$ 2000) accompanied with the large white-light flares, 
(2) some weak H$\alpha$/X-ray flares are not accompanied with white-light emissions, and  
(3) the non-flaring emissions show clear rotational modulations  in X-ray and H$\alpha$ intensity in the same phase.
To understand these observational features, one-dimensional  hydrodynamic  flare simulations are performed by using the RADYN code.
As a result of simulations,  we found the simulated H$\alpha$ line profiles with hard and high-energy non-thermal electron beams are consistent  with that of the initial phase line profiles of the superflares, while those with more soft- and/or weak-energy beam are consistent with those in decay phases, indicating the changes in the energy fluxes injected to the lower atmosphere.
Also, we found that the relation between optical continuum and H$\alpha$ intensity is nonlinear, which can be one cause of the non-white-light flares.
 The flare energy budget exhibits diversity in the observations and models, and more observations of stellar flares are necessary for constraining the occurrence of various emission line phenomena in stellar flares. 

\end{abstract}

\section{Introduction}\label{sec:intro}

%{\bf\color{red} \color{black}}

%Introduction of solar and stellar flare
Solar flares are abrupt brightenings on the solar surface.
During flares, magnetic energy stored around sunspots is believed to be converted to kinetic and thermal energies through the magnetic reconnection in the corona (see, \cite{1981sfmh.book.....P}; \cite{2011LRSP....8....6S} and reference therein). 
In the standard scenario, the released energies are transported from the corona to the lower atmosphere by non-thermal high-energy particles and thermal conduction. 
The energy injection causes chromospheric evaporations and chromospheric condensations, producing bright coronal and chromospheric emissions, respectively.
In this context, these chromospheric/coronal emissions have information on the accelerated particles in the reconnection site, which can give us a clue to understanding the unknown acceleration mechanism of the non-thermal particles.

As expected from a solar analogy, stellar flares are often observed in radio, visible, and X-ray ranges similar to solar flares. 
In particular, magnetically-active stars, such as young T-tauri stars (e.g., \cite{1996PASJ...48L..87K}, \cite{2010ARA&A..48..241B}) and M-type stars (e.g., \cite{1991ApJ...378..725H}; \cite{2013ApJS..207...15K}), often show large flares, called superflares.
The superflares release much larger total energies (10$^{33}$ -- 10$^{38}$ erg) than the maximum solar flares ($\sim$10$^{32}$ erg).
This kind of extreme event on the stars has been getting more and more attention in terms of the exo-planet habitability around active young stars \citep{2010AsBio..10..751S,2016NatGe...9..452A,2017ApJ...848...41L} and a possible extreme event on the Sun \citep{2013A&A...549A..66A,2013PASJ...65...49S,2017ApJ...850L..31H}.

%History of observation of the Balmer lines on M-dwarf flares

As the solar flare dynamics have been well-understood thanks to the multi-wavelength observations of solar flare, the understanding of the large stellar flares is expected to be deepened by them.
%However, {\bf \color{red} the number of stellar flares successfully detected by multi-wavelength observations is still not so large compared to solar flares,  and
More samples are required to reveal the universality and diversity of solar and stellar flares. 
Magnetically active M dwarfs are one of the best targets for the flare monitoring, whose flares are observed from X-ray to radio.
Particularly, stellar flares produce greatly enhanced emission in chromospheric lines, such as the hydrogen Balmer series, Ca II H and K, which are observable from the ground.
The hydrogen lines tend to have a relatively fast rise phase, but the peak is often delayed compared to the continuum emission (\cite{1982ApJ...252..239K}; \cite{1991ApJ...378..725H}).
The radiated energy in hydrogen lines is relatively small compared to the continuum \citep{1991ApJ...378..725H}.
The Balmer line broadening up to 20 {\AA} has been observed during stellar flares \citep{1991ApJ...378..725H}, which is interpreted as the non-thermal broadening or Stark (pressure) broadening.
Recent numerical simulation shows that the broadenings of the higher-order Balmer lines (e.g. H$\gamma$) are good tools to estimate the chromospheric density, which can be a clue to the injected accelerated particles \citep{2006ApJ...644..484A,2006PASP..118..227P,2017ApJ...837..125K}.
However, in almost all studies, the temporal evolution of the Balmer line widths has not been well investigated although the flaring atmosphere dramatically changes during flares.
Moreover, the energy budget for each wavelength is not confirmed for stellar flares, although it is known that there is diversity even in solar flares.

%This work & Section Introduction
In this paper, we report the optical and X-ray monitoring observation of an M-dwarf flare star AD Leo during 8.5 nights by Seimei-OISTER campaign to reveal the features of stellar flares.
In this campaign, we mainly used a low-resolution spectrograph on the 3.8-m Seimei Telescope (\cite{Kurita2020}).
We also conducted optical spectroscopy and photometric observations with the help of the OISTER (Optical and Infrared Synergetic Telescopes for Education and Research \footnote[1]{http://oister.kwasan.kyoto-u.ac.jp/}) program and  with the SCAT at the Chuo University. 
We also obtained the X-ray monitoring data from NICER (Neutron Star Interior Composition Explorer) during this observational period.
 In Section \ref{sec:2}, we review observations and analyses. In Section \ref{sec:3}, we introduce features of the observed stellar flares. In Section \ref{sec:4}, we show the rotational modulations of the AD Leo. In Section \ref{sec:5}, we perform one-dimensional hydrodynamic simulations of stellar flares to understand the flare properties. Finally, we discuss the observations and numerical simulations in Section \ref{sec:6}.

\section{OBSERVATIONS AND DATA REDUCTION} \label{sec:2}
\subsection{Target Star} 
In 2019, we carried out large campaign monitoring observations on the nearby M dwarf AD Leo (GJ 388).
AD Leo is classified to a dMe 3.5 star \citep{2009ApJ...699..649S}, whose distance from the Earth is about 4.9 pc.
Frequent stellar flares have been observed on AD Leo with the several wavelength ranges \citep{1995ApJ...453..464H,2003ApJ...597..535H,2013ApJS..207...15K}, and an extremely large superflare was also observed \citep{1991ApJ...378..725H}.
The flare occurrence frequency is reported to have a power-law distributions, causing 0.76 flares per day \citep{1984ApJS...54..375P}.

%\begin{longtable*}{llccc}
%\tablenum{1}
%\caption[]{Observing Log\label{tab:obslog}}
\begin{table*}[t]
\tbl{Observing Log}{%
\begin{tabular}{llccc}
%\tablewidth{0pt}
\hline
Telescope/Instrument & UT Date (JD) & Time & Exp Time & Flares {\#} \\
(Data type) &  &  (hr) & (s) &  \\
\hline
\hline
\textsf{Spectroscopy}&&&& \\
\hline
%\colhead{Messier} & \colhead{NGC/IC} & \nocolhead{Common} & \colhead{Object} &
%\multicolumn2c{Distance} & \colhead{} & \colhead{V} \\
%\colhead{Number} & \colhead{Number} & \nocolhead{Name} & \colhead{Type} &
%\multicolumn2c{(kpc)} & \colhead{Constellation} & \colhead{(mag)}
%\decimalcolnumbers
%\startdata
3.8m Seimei/KOOLS-IFU & 2019 Mar 22 (2458565) & 2.6 & 60 & {\#}1 \\
(5000-8000 {\AA}; R$\sim$2000)  & 2019 Mar 23 (2458566) & 4.6, 1 & 30 & -  \\
 & 2019 Mar 24 (2458567) &   4.1 & 60 & {\#}2, 3 \\
 & 2019 Mar 25 (2458568) &   2.5 & 30, 60 & -  \\
 & 2019 Mar 26 (2458569) &   7.3 & 30 & {\#}4, 5 \\
 & 2019 Mar 27 (2458570) &   5.1 & 30 & {\#}6 \\
%& 2019 Mar 28 (24585XX) &   2, 2 & 30 & SF2, SF3 \\
 & 2019 Apr 12 (2458586) &   5.9 & 30 & {\#}7, 8, 9, 10 \\
\hline
2m Nayuta/MALLS (OISTER) & 2019 Mar 24 (2458567)  &  5 & 120 & {\#}2, 3 \\
(6350-6800 {\AA}; R$\sim$10000) & 2019 Mar 26 (2458569)  &  5 & 120 & {\#}4, 5 \\
\hline
36cm SCAT & 2019 Mar 23 (2458566) &  2.5 & 600 & \\
(3520-8040 {\AA}; R$\sim$600; & 2019 Mar 24 (2458567) &  4.5 & 600 & {\#}2, 3 \\
 covering H$\alpha$, H$\beta$, H$\gamma$, and H$\delta$)  & 2019 Mar 26 (2458569) &  5.5 & 600 & {\#}4, 5 \\
 & 2019 Mar 27 (2458570) &  2.5 & 600 & \\
\hline
\textsf{Photometry}&&&& \\
\hline
 50cm MITSuME  (OISTER) & 2019 Mar 22 (2458565) &  6.9 & 5 & {\#}1 \\
(g'/Rc/Ic-band photometry)$^\dagger$ & 2019 Mar 23 (2458565)  &  6.0 & 5 & \\
 & 2019 Mar 24 (2458567) &  6.8 & 5 & -$^{*}$ \\
 & 2019 Mar 25 (2458568) &  4.8 & 5 & \\
 & 2019 Mar 26 (2458569) &  7.0 & 5 & -$^{*}$ \\
% & 2019 Mar 27 (24585XX) &   2, 2 & 5 & SF2, SF3 \\
 & 2019 Mar 28 (2458571) &  7.0 & 5 &  \\
 & 2019 Apr 11 (2458585) &  4.5 & 5 &  \\
 & 2019 Apr 12 (2458586) &   4.5 & 5 & -$^{*}$ \\
 & 2019 Apr 13 (2458587) &  4.6 & 5 &  \\ 
\hline
40cm KU Telescope  (OISTER) & 2019 Apr 12 (2458586) &  5 & 10 & {\#}7, 9, 10 \\
(B-band photometry)$^\dagger$ &  &   &  & \\
\hline
\textsf{X-ray}&&&& \\
\hline
ISS/NICER & 2019 Mar 22-28 (2458565-71) & $\sim$0.5 × 26 & - & {\#}4 \\
(0.2-12 keV X-ray) & 2019 Apr 11-13 (2458585-87) & $\sim$0.5 × 10 & - & {\#}8, 11, 12 \\
\hline
%\multicolumn{5}{l}{$^{*}$ There were observations when flares were detected by Seimei Telescope and other telescope, but no flares are detected by MITSuME. }
%\endlastfoot
%$^{*}$ There were observations when flares were detected by Seimei Telescope and other telescope, but no flares are detected by MITSuME.
%\hline
%\end{longtable*}
\end{tabular}}\label{table:extramath}
\begin{tabnote}
\hangindent6pt\noindent
\hbox to6pt{\footnotemark[$*$]\hss}\unskip% 
 There were observations when flares were detected by Seimei Telescope and other telescope, but no flares are detected by MITSuME. $^\dagger$ $g'$-, $R_c$-, $I_c$-, $B$-band filter is broad-band (full width$\sim$ 1000 {\AA}) one whose central wavelength are 4858, 6588, 8060, 4448{\AA}, respectively. 
\end{tabnote}
\end{table*}

\subsection{Spectral Data}
We mainly used the Seimei Telescope located at Okayama Observatory, Japan, for spectroscopic data.
The Seimei Telescope is 3.8 m optical and infrared telescope (\cite{Kurita2020}).
We used the KOOLS-IFU instrument \citep{2019PASJ...71..102M}, which is a low-resolution spectrograph (KOOLS) with an optical-fiber integral field unit (IFU), on the Nasmyth focus.
The grism we used covers 6150 to 7930 {\AA}, and the spectral resolution (R) is $\sim$ 2,000.
We conducted the 8.5 nights of spectroscopic monitoring observation of AD Leo with Seimei Telescope/KOOLS-IFU during March to April 2019 (Table 1).
The time resolutions are 42 or 72 seconds including the 12-sec read-out time to achieve the signal to noise $\sim$ 100. 
The spectroscopic data of the KOOLS-IFU are two-dimensional spectroscopic data, and we use only the fiber array where stellar integrated brightness is more than 50 \% than that of the maximum fiber.
Data reduction was done using the package of the IRAF\footnote[2]{IRAF and PyRAF are distributed by the National Optical Astronomy Observatories, which are operated by the Association of Universities for Research in Astronomy, Inc., under cooperate agreement with the National Science Foundation.} and PyRAF$^2$ software and the data reduction packages developed by \citet{2019PASJ...71..102M}. \footnote[3]{http://www.kusastro.kyoto-u.ac.jp/~kazuya/p-kools/reduction-201806/index.html}

%During this observational period of Seimei Telescope, we also conducted monitoring observations of the Balmer lines of AD Leo with SCAT Telescope (Tsuboi et al. in prep).
%SCAT Telescope is the 36-cm telescope located at Chuo Univerisity, Japan, and it covers 3520 - 8040 {\AA} with the spectral resolution of XX.
%About 600-sec exposure was required to get the signal to noise of $>$ 100.

During this observational period of Seimei Telescope, we also conducted monitoring observations of the Balmer lines of AD Leo with optical telescope SCAT (Spectroscopic Chuo-university Astronomical Telescope). SCAT is mounted on a building in Korakuen campus of Chuo University in Japan. It consists of an MEADE 36 cm diameter telescope and an ATIK 460EX CCD camera with an Shelyak Alpy 600 spectrometer. The spectrometer covers 3700 to 7500 {\AA}, and the spectral resolution, R, is 600. About 600-sec exposure was required to get the signal to noise of $>$ 100. We executed the data reduction using the twodspec package of the IRAF software in the standard manner (dark subtraction, flat fielding, spectral extraction, sky subtraction, and wavelength calibration).

 In the OISTER program,  the spectroscopic observations were carried out with the Nayuta 2 m telescope at the Nishi-Harima Astronomical
Observatory for two days (Table 1).
The MALLS (Medium And Low-dispersion Long-slit Spectrograph) was used with a resolving power (R) of $\sim$10000 at 6500 {\AA}, covering 6350 - 6800 {\AA}.
We aimed to use this instrument to detect line asymmetries of the Balmer lines (e.g., \cite{2018PASJ...70...62H}),  but the changes in the H$\alpha$ profiles were too small, and the significant line asymmetries were not detected. 

\subsection{Photometric Data}
 In the OISTER program,  time-resolved photometry was performed during this period by using MITSuME 50 cm telescope at Okayama Observatory and  the 40-cm telescope at Kyoto University. 
MITSuME 50cm-telescope can acquire $g'$, $R_{\rm C}$, and $I_{\rm C}$-band images simultaneously by using two dichroic mirrors and three CCD cameras \citep{2005NCimC..28..755K}.
We described the observational log of MITSuME in Table 1.
Note that although the location is the same as that of Seimei Telescope, the photometry has better sensitivity than the spectroscopy of Seimei.
The CCDs of MITSuME have deteriorated recently, and the photometric sensitivity has become worse if we divide the images by flat flames.
Therefore, most flares except for one large superflare could not be detected by MITSuME photometry even in $g'$-band where flare amplitude is expected to be the largest among the three bands of MITSuME.
 Also, B-band photometric observations on AD Leo was conducted by the 40-cm telescope at Kyoto University only on April 12th, and the data are shown in  Appendix \ref{app:1}. 

\subsection{X-ray Data}

NASA's Neutron Star Interior Composition Explorer (NICER; \cite{2016SPIE.9905E..1HG}) has conducted the monitoring observations on AD Leo during this period.
NICER is the soft X-ray instrument onboard the International Space Station, and observed AD Leo for about 1 ksec for each orbital period of ISS (about 90 minutes).
The observation has been carried out for several times during each night.
NICER is not an imaging instrument, so background spectra must be subtracted to get the stellar spectra.
The data were processed using NICER software version 2019-10-30, which can estimate the background spectra at a given NICER observational orbit.\footnote[4]{https://heasarc.gsfc.nasa.gov/docs/nicer/tools/nicer{\_}bkg{\_}est{\_}tools.html}

In making the light curves, we used 0.5-8 keV band corresponding channels 50 through 800.
Below channel $\sim$ 50 and above $\sim$ 800, there is optical contamination due to the ambient light.
For two flares clearly detected by NICER (Flare \#4 and \#12), we also analyzed X-ray spectra in flare phases.
The integrated times are indicated by the error bars of the derived emission measure (EM) and temperature as in Figure \ref{fig:sf4}.
We fitted the pre-flare subtracted X-ray spectra (0.5-8.0 keV) with a simple thin-thermal model of single-temperature plasma where the abundance ratios of heavy elements are fixed to the solar values (e.g., \cite{2016PASJ...68...90T}).
We derived the emission measures, temperatures, and radiation flux by using the $apec$ models in \textit{XSPEC} installed in HEASoft \footnote[5]{https://heasarc.gsfc.nasa.gov/xanadu/xspec/}, and the parameters are summarized in Table \ref{table:xspec}.

\begin{table*}
\tbl{X-ray spectral best-fit parameters for flare {\#}4}{%
\begin{tabular}{lccc}  
\hline\noalign{\vskip3pt} 
Parameters & time 1 & time 2  & time 3  \\
\hline\noalign{\vskip3pt} 
$N_{\rm H}$ [$10^{20}$ $\rm cm^{-2}$]       &  3.48 &  8.50 &  5.23   \\
$kT$ [keV]       &  2.62  & 1.31  &   1.27     \\
norm       &  8.14 $\times$ 10$^{-2}$  & 5.07 $\times$ 10$^{-2}$   &    3.10 $\times$ 10$^{-2}$    \\
\hline\noalign{\vskip3pt} 
%qi-squared    &   &   &        \\
\hline\noalign{\vskip3pt} 
\end{tabular}}\label{table:xspec}
\begin{tabnote}
\hangindent6pt\noindent
%\hbox to6pt{\footnotemark[$*$]\hss}\unskip% 
% Symbols provided by the standard \LaTeX{} system such as $\cong$, $\approx$ are available. 
% If the \textsf{amssymb} package is available, then the \textsf{useamsfonts} class option enables 
% to use the symbols defined by \textsf{amssymb} package. 
% (Also note that this document is \textit{not} an instruction for \LaTeX{} itself, 
% we omit a list for those symbols.)
\end{tabnote}
\end{table*}

%%NICER subset we use%%%
%For this analysis we consider only the detector channel12 subset [25, 300)—meaning channels 25 through 299
%inclusive—nominally corresponding to 0.25\UTF{2013}3 keV. Below channel 25, there is increased “optical loading” contamination (electronic noise due to ambient light), and there is greater uncertainty in the detector readout triggering
%efficiency for valid X-ray events. Above channel 300 the soft thermal emission from PSR J0030+0451 becomes negligible relative to the non-source background.

\subsection{Emission Line/Continuum Fluxes}
Emission fluxes were calculated for the hydrogen Balmer lines (H$\alpha$, H$\beta$, H$\gamma$, H$\delta$), and the He I line 6678.15{\AA}, and the $g'$, $R_{\rm C}$, and $I_{\rm C}$-band continuum.
For the emission lines, the flux ($F_{\rm line}$) is calculated from equivalent width (EW) and the local continuum enhancement levels ($F_{\rm flare}/F_{\rm pre-flare}$),  and the local continuum enhance level is calculated based on $g'$-band and $R_{\rm C}$-band (c.f. the Appendix of \cite{1991ApJ...378..725H}).
First, the synthetic $g'$-band flux, $R_{\rm C}$-band flux, and local continuum flux ($F_{\rm local-cont.}$) at each emission line in quiescence is calculated based on the flux-calibrated AD Leo spectra taken by SCAT.
Second, the local-continuum enhancement levels at each line (i.e. $F_{\rm flare}/F_{\rm pre-flare}$) is calculated based on the $g'$- and $R_{\rm C}$-band enhancements level obtained from photometry.
Finally, the line emission flux is calculated from the equivalent width and local-continuum enhancements level (i.e., $F_{\rm line}$ = EW $\times$ $F_{\rm flare}/F_{\rm pre-flare}$ $\times$ $F_{\rm local-cont.}$). 
H$\alpha$ and  He I line 6678.15{\AA} refer to the $R_{\rm C}$-band flux, and the others do to the $g'$-band flux.
The fluxes of line emissions can have errors because of the contamination of line emissions on the broad-band continuum fluxes.
However, for example, as for the flare \#1, the effect would be less effective because the enhancement of equivalent width was 10 {\AA} at most while the continuum bands have $>$ 100 \% enhancement in $\sim$ 1000 {\AA} bandwidth.

\section{Flare Atlas: Light curves and Spectra} \label{sec:3}
\subsection{Observational Summary} \label{sec:3.1}

\begin{figure*}
\begin{center}
\includegraphics[scale=0.5]{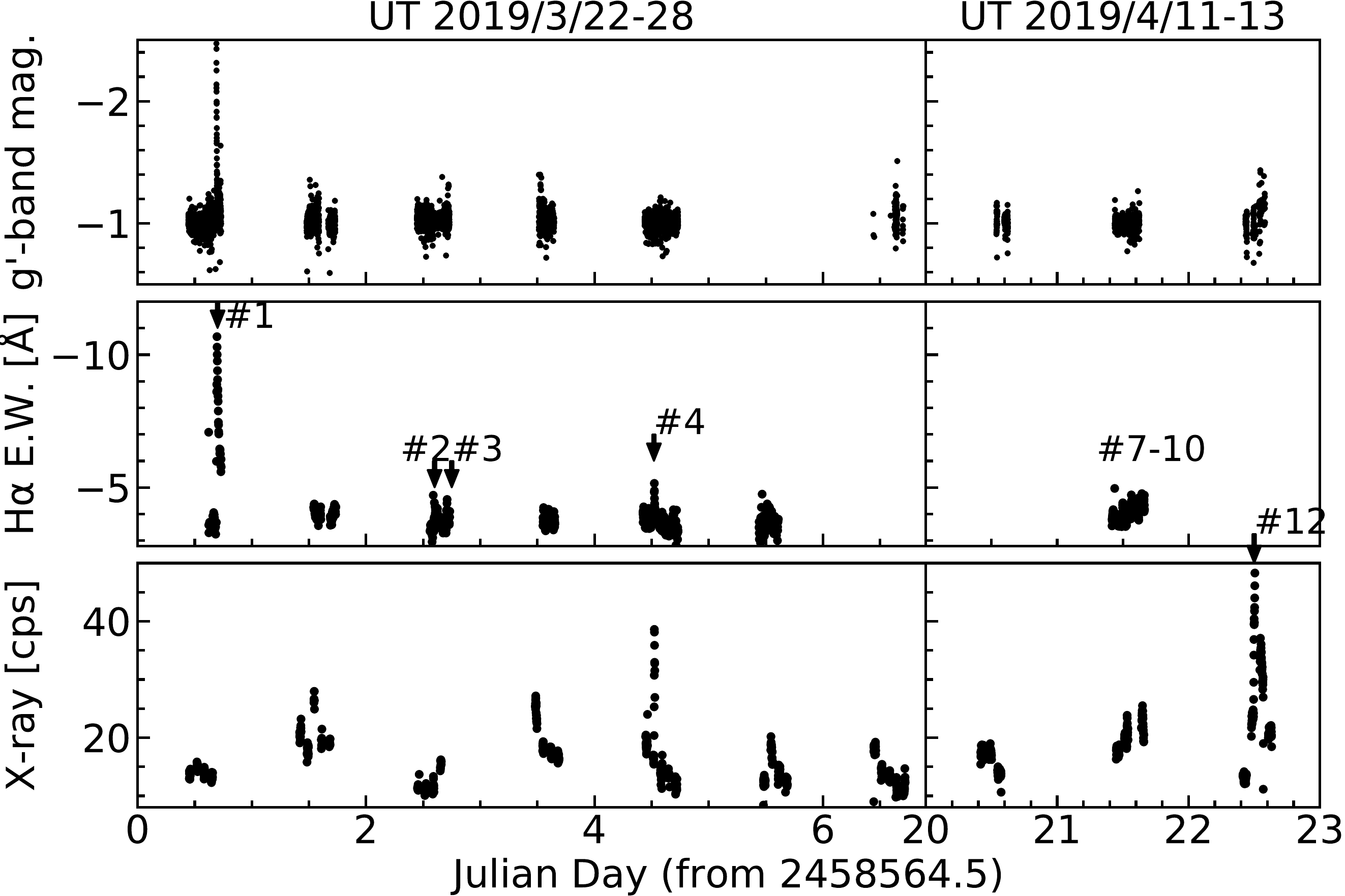}
\end{center}
\caption{ Overall light curves of AD Leo during this campaign observation. Top: the light curve in $g'$-band magnitude. Middle: the H$\alpha$ E.W.. Bottom: X-ray count rates [cps; count per sec] in 0.5-8 keV. }
\label{fig:lcall}
\end{figure*}

We carried out the monitoring observations on AD Leo for 8.5 days, and the clear-sky ratio was about 50 \%.
 Figure \ref{fig:lcall} indicates the overall light curve during this campaign. 
12 flares were detected by eye  mainly with the H$\alpha$ monitoring with Seimei/KOOLS-IFU (see Table 1) although there could be a larger amount of small flares which could not be identified by eye.
The Balmer lines show emission even in quiescence, indicating very high atmospheric heating.
The H$\alpha$ equivalent width in quiescence is about -3.5 {\AA}, and the enhancement during flares are typically 1-1.5 {\AA}.
Only one flare (flare \#1) shows very high enhancement of H$\alpha$ $\sim$ 10 {\AA}.
The number of flares detected by Balmer lines is 10.
Even though the simultaneous photometry is limited, four of them are clearly detected by optical photometry, while five of them did not clearly show the white-light emissions  (one of them has no photometry). 
Four of them are also detected by higher resolution spectroscopy by Nayuta/MALLS, but are too weak to identify clear line asymmetry like \citet{2018PASJ...70...62H}.
Two of them are detected by NICER X-ray detector.
The count rates of 0.5-8.0 keV in quiescence is about 18 counts per second.
Besides, two additional flares are detected by NICER X-ray although there are no clear H$\alpha$ observations (flare \#11, \#12).
In the following sections, we show the typical and prominent stellar flares detected (flare \#1, \#2, \#3, and \#4), and the other all flares are shown in the Appendix.

\subsection{Flare \#1: A Superflare Showing Large Line Broadening} \label{sec:3.2}

\begin{figure}
\begin{center}
\includegraphics[scale=0.4]{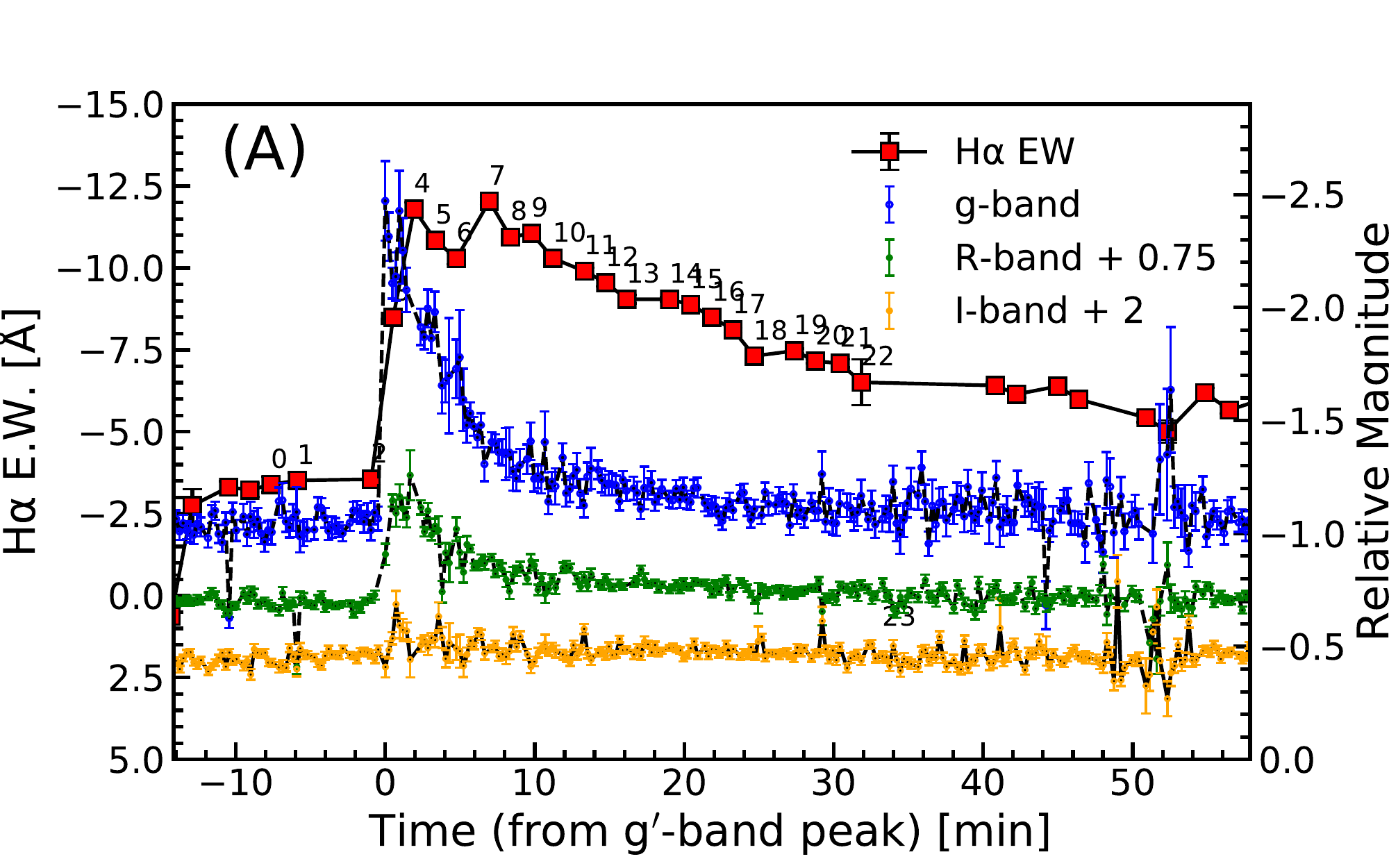}
\includegraphics[scale=0.4]{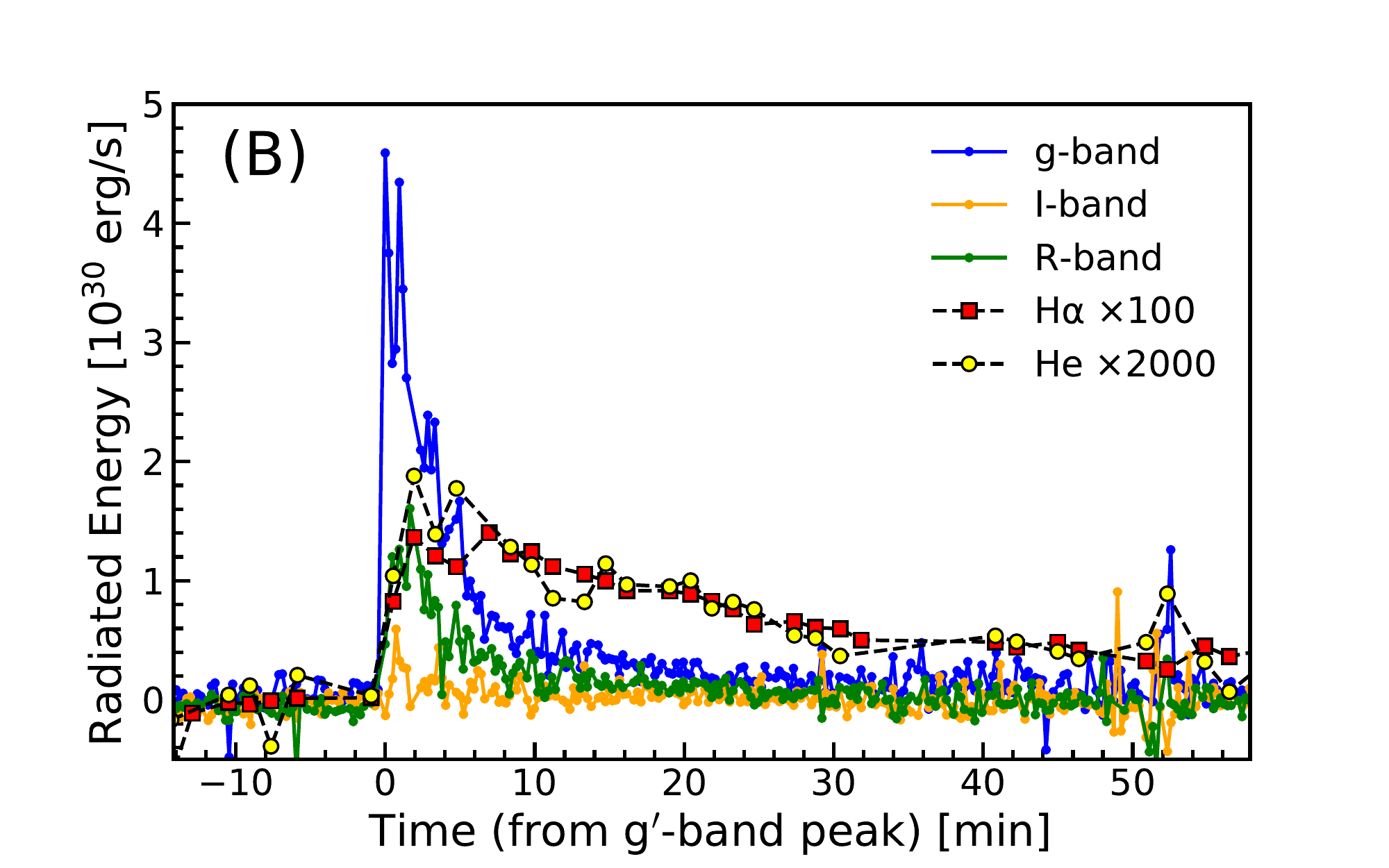}
\end{center}
\caption{ (A) Light curve of the flare \#1 observed by Seimei Telescope/KOOLS-IFU (H$\alpha$) and MITSuME ($g^{'}$, $R_{\rm c}$, $I_{\rm c}$).  The equivalent width is corrected by the continuum enhancements  (B) Light curve of flare \#1 in the unit of erg$\cdot$s$^{-1}$. }
\label{fig:lc1}
\end{figure}

Figure \ref{fig:lc1} shows the light curve of the flare \#1 observed by Seimei spectroscopy and MITSuME photometry. 
Unfortunately, there were no X-ray observations during this flare.
The panel (A) shows that the equivalent width of H$\alpha$ becomes -12 from -3.5 {\AA}.
The $g'$-band continuum becomes four times brighter than the quiescence, and the contrast is largest among the three filters.
This can indicate very blue spectra of white-light continuum emissions,  but M dwarfs are also very red so a flat spectrum leads to larger flux enhancements as well. 
The panel (B) shows the temporal evolution of the radiated flux for each wavelength.
The energies radiated in the continuum bands are much larger than the line emissions.
The total radiated energies (and ratios relative to the g-band energy) in $g'$-band, $R_{\rm C}$-band, $I_{\rm C}$-band continuum, H$\alpha$, and He lines are calculated to be 1.4$\times10^{33}$ erg, 4.7$\times10^{32}$ erg (0.34), 7.0$\times10^{31}$ erg (0.05), 2.5$\times10^{31}$ erg (0.018), and 1.3$\times10^{30}$ erg (0.0093), respectively, and the flare is classified to be a superflare  (a flare with the total energy of more than 10$^{33}$ erg $\sim$ ten times larger energy than the largest scale of solar flares; \cite{2012Natur.485..478M}).
Here, to calculate the flare energies, the flare fluxes in continuum and line emission in Figure \ref{fig:lc1} (B) were time-integrated between -2.6 min and 36 min and between -2.6 min and 58 min, respectively.
Because the observations finished before the H$\alpha$ and He line flare emissions completely decayed, the energies of line emissions would be underestimated to some extent. 
%The flare is classified to be a superflare.
The duration of the H$\alpha$ flare is more than one hour, while those of white-light flares are about 15 minutes.
The continuum fluxes have shorter durations than the chromospheric line emission, which can be an indication of the  Neupert  effect in the case of solar flares \citep{1968ApJ...153L..59N}.
 The color temperature of the white-light emission during the flare is calculated to be typically 14,000 $^{+17,000}_{-8,000}$ K if we assume the black-body radiation for $g'$-band and $R$-band fluxes. 
Note that the continuum flux ratio was very noisy during the flare, so the error bar of the emission temperature is very large. 
The temporal evolution of the white-light emission temperature is therefore not significantly found. 
 However, it is reported that broad-band continua, especially $g'$-band, could be affected by emission lines (e.g., \cite{2019ApJ...878..135K}), so we need to be careful about the interpretations of the emission temperatures derived here. 

\begin{figure}
\begin{center}
\includegraphics[scale=0.4]{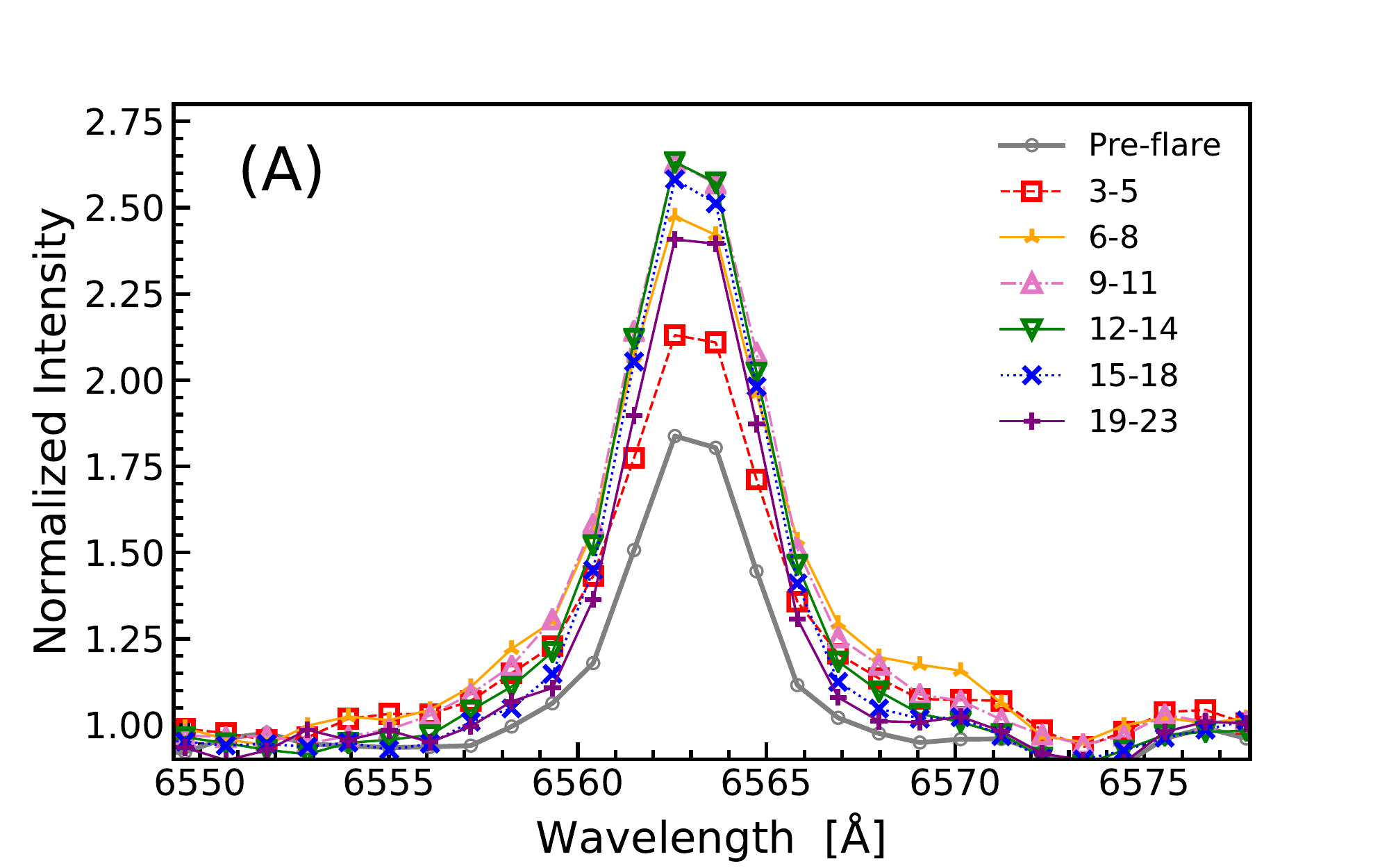}
\includegraphics[scale=0.4]{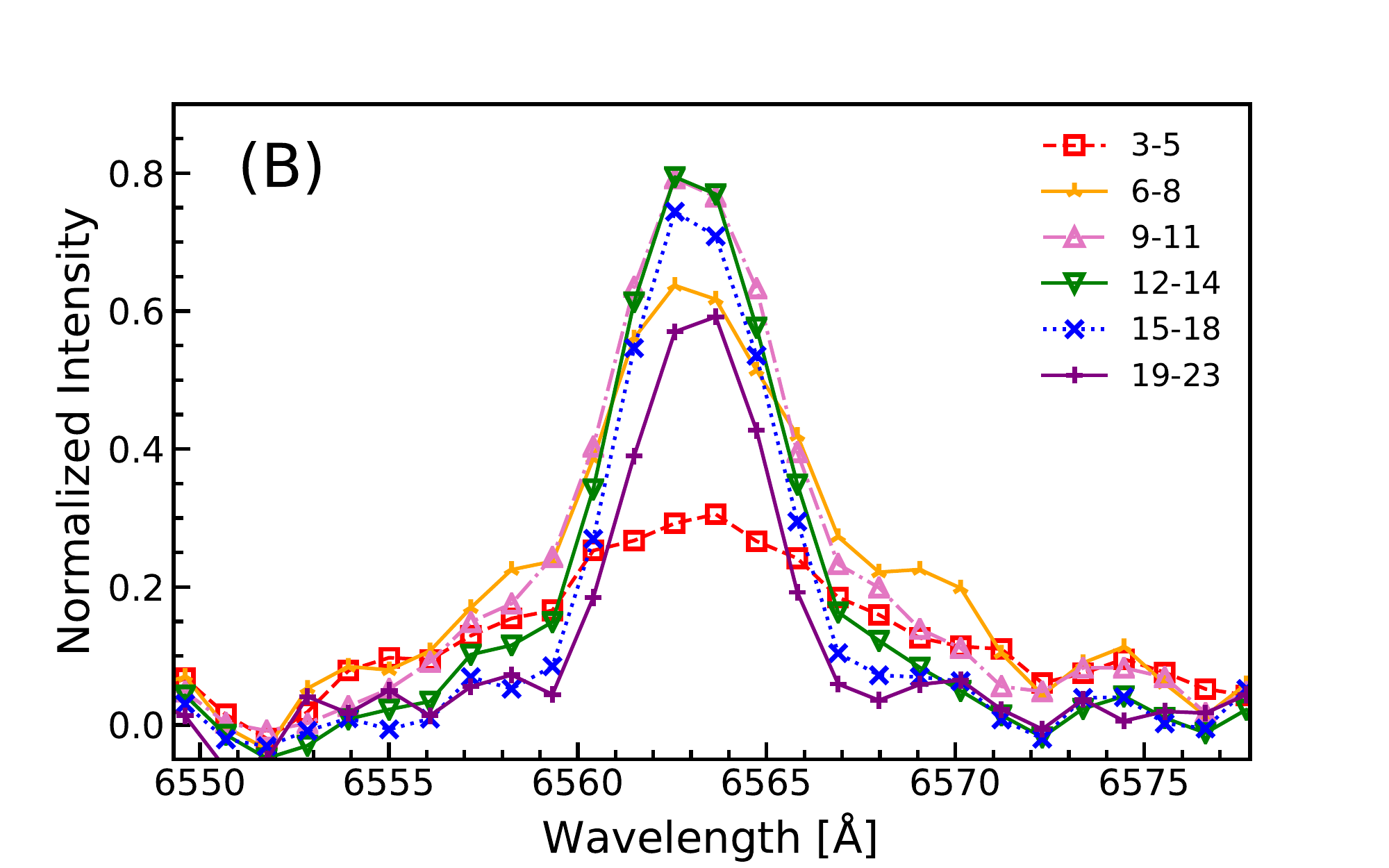}
\end{center}
\caption{(A) H$\alpha$ spectra during the flare \#1. The numbers described in the legend is the same ones shown in the Figure \ref{fig:lc1}. We combined three to five spectra to make the spectra in the panel (A). (B) Pre-flare subtracted spectra of the flare \#1. } 
\label{fig:flare1spectra}
\end{figure}

\begin{figure}
\begin{center}
\includegraphics[scale=0.4]{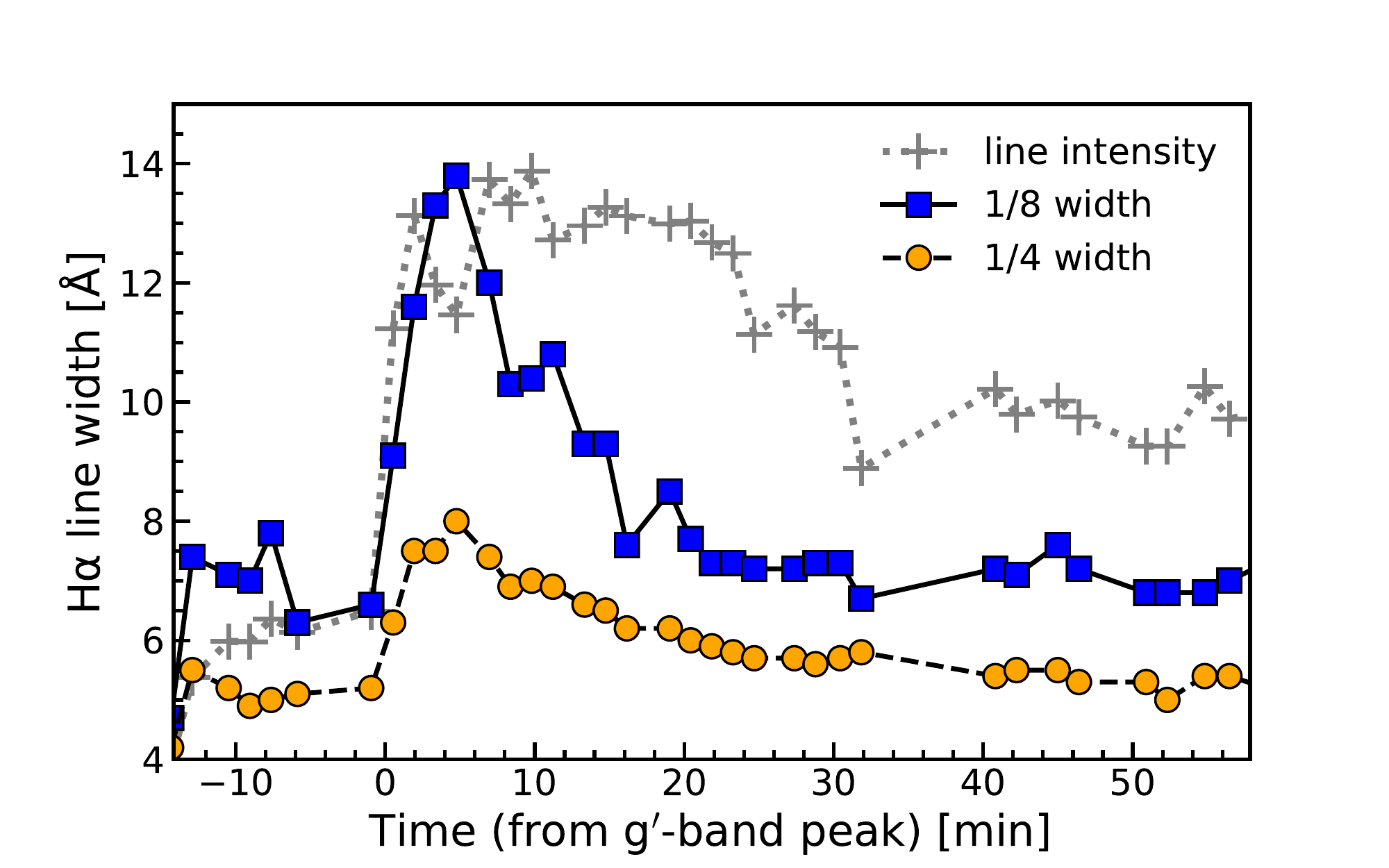}
\end{center}
\caption{Temporal evolution of the H$\alpha$  line width and line intensity of the flare \#1. Open squares and open circles indicates the line width where the line intensity is 1/8 and 1/4 of the peak intensity, respectively. Gray crosses are the scaled line peak intensity.}
\label{fig:wingint}
\end{figure}

Figure \ref{fig:flare1spectra} shows the temporal evolution of the low-resolution H$\alpha$ spectra during this flares.
The panel (A) is the spectra  normalized by the continuum level, and the panel (B) is the pre-flare subtracted spectra.
First, we could not find any line asymmetry during this flare, although the blue and red asymmetries are frequently observed during not only stellar flares but also solar flares  (e.g., \cite{1984SoPh...93..105I}; \cite{2018PASJ...70..100T} ;\cite{2018PASJ...70...62H}; \cite{2020arXiv200306163M}). 
Second, we found significant line-wing broadenings of H$\alpha$ line during the flare, as you can see in Figure \ref{fig:flare1spectra} (B).
The line broadening is prominent especially in the initial phase of the flare, but it is not prominent in the later decay phase.
Figure \ref{fig:wingint} shows the temporal evolution of the line width and line peak intensity.
As you can see, both line width and intensity largely increase in the initial phase of the flare when the white-light emissions are seen.
In the decay phase, the line width dramatically decreases while the line peak intensity does not largely change.

\subsection{Flare \#2 (and \#3): A Flare Showing clear Balmer-line Decay} \label{sec:3.3}

\begin{figure}[htbp]
\begin{center}
\includegraphics[scale=0.4]{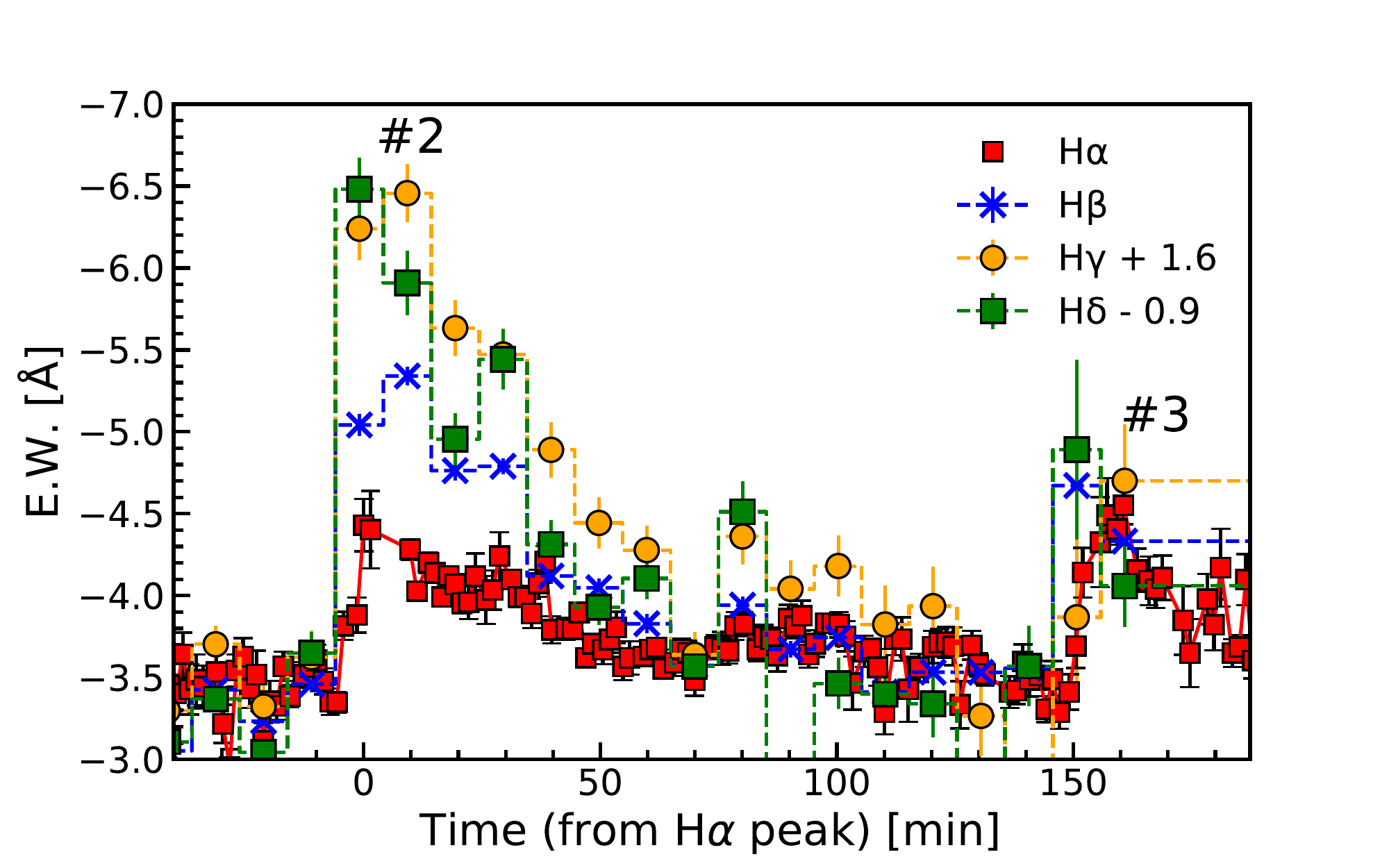}
\end{center}
\caption{Equivalent width (E.W.) of Balmer lines during the flare \#2 and \#3 observed by Seimei Telescope and SCAT. The relative values are corrected and the pre-flare levels are set to be the same value. }
\label{fig:lc23}
\end{figure}

Figure \ref{fig:lc23} shows the light curve of flare \#2 and \#3, observed by Seimei/KOOLS-IFU and SCAT.
There were photometric observations by MITSuME during this flare, but no significant white-light enhancement can be seen. 
The equivalent width change of the higher level Balmer line  (e.g. H$\delta$ and H$\gamma$) is larger than the lower level (e.g. H$\alpha$),  which would be because there is a lower continuum in the blue. 
However, the decay timescale for each line is quite similar to each other.
%On the other hand, as in the panel (B), the light curves and durations look similar with each other.
The decay timescale for each Balmer lines is not easy to interpret because temperature, density, and the difference in opacity for each line contribute to it.

\subsection{Flare \#4: A Small Flare Observed by the All Instruments} \label{sec:3.4}

\begin{figure}
\begin{center}
\includegraphics[scale=0.4]{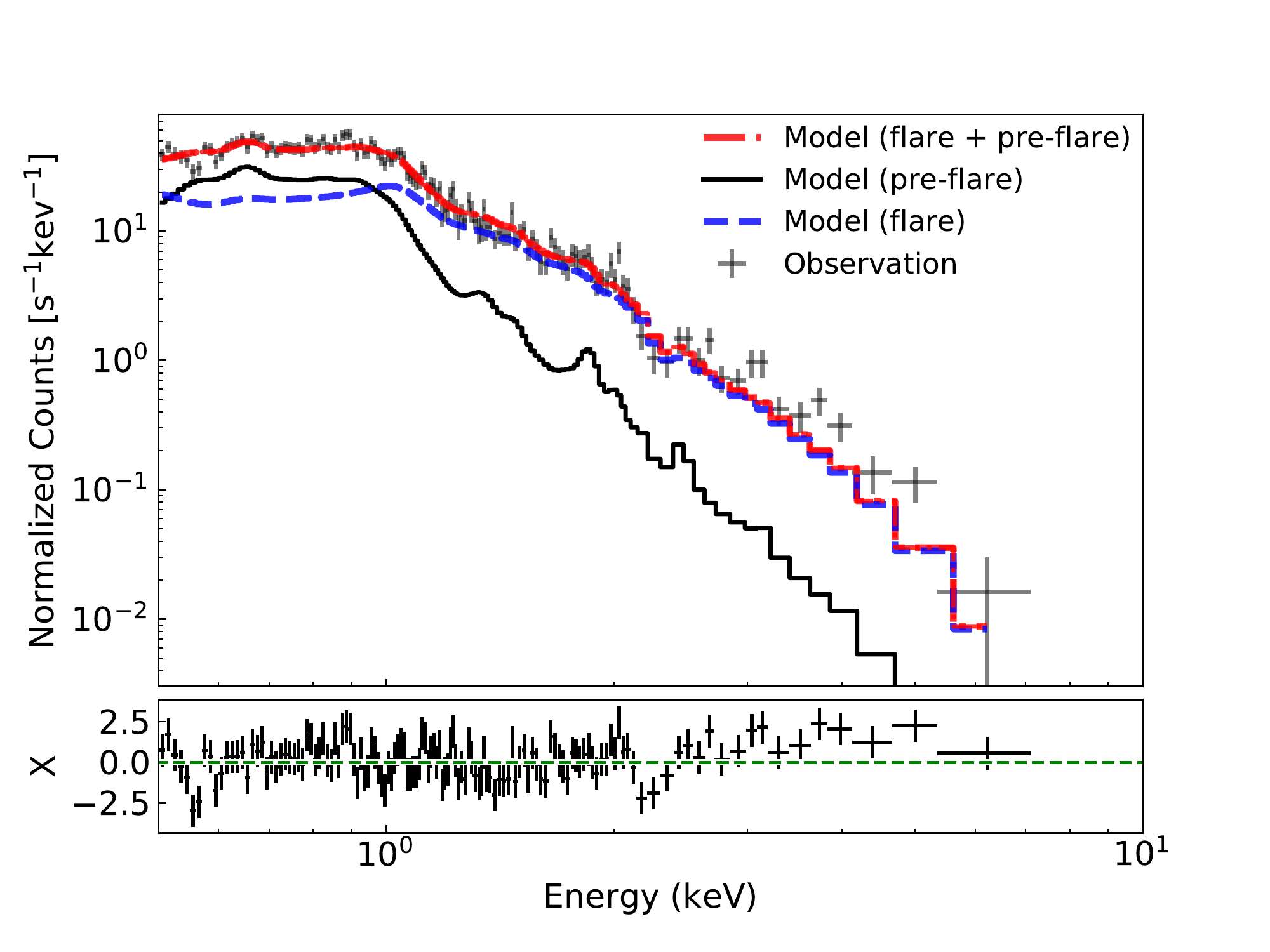}
\end{center}
\caption{X-ray spectra (0.5 - 10 keV) at peak of flare {\#}4. The gray crosses indicate the X-ray spectrum observed by NICER.  The red dash-dotted line indicates the model spectra obtained by fitting the observation with a simple thin-thermal model ($apec$ model in $XSPEC$). The black solid line is the pre-flare spectra, and the blue dashed line indicate the flare-only spectra (i.e. the red line minus the black line).   As a result of the model fitting, the reduced chi-squared is 1.23.  The bottom panel indicates values of  X = (data - model)/error (normalized by one sigma)  for each bin.}
\label{fig:xspec4}
\end{figure}

\begin{figure}
\begin{center}
\includegraphics[scale=0.4]{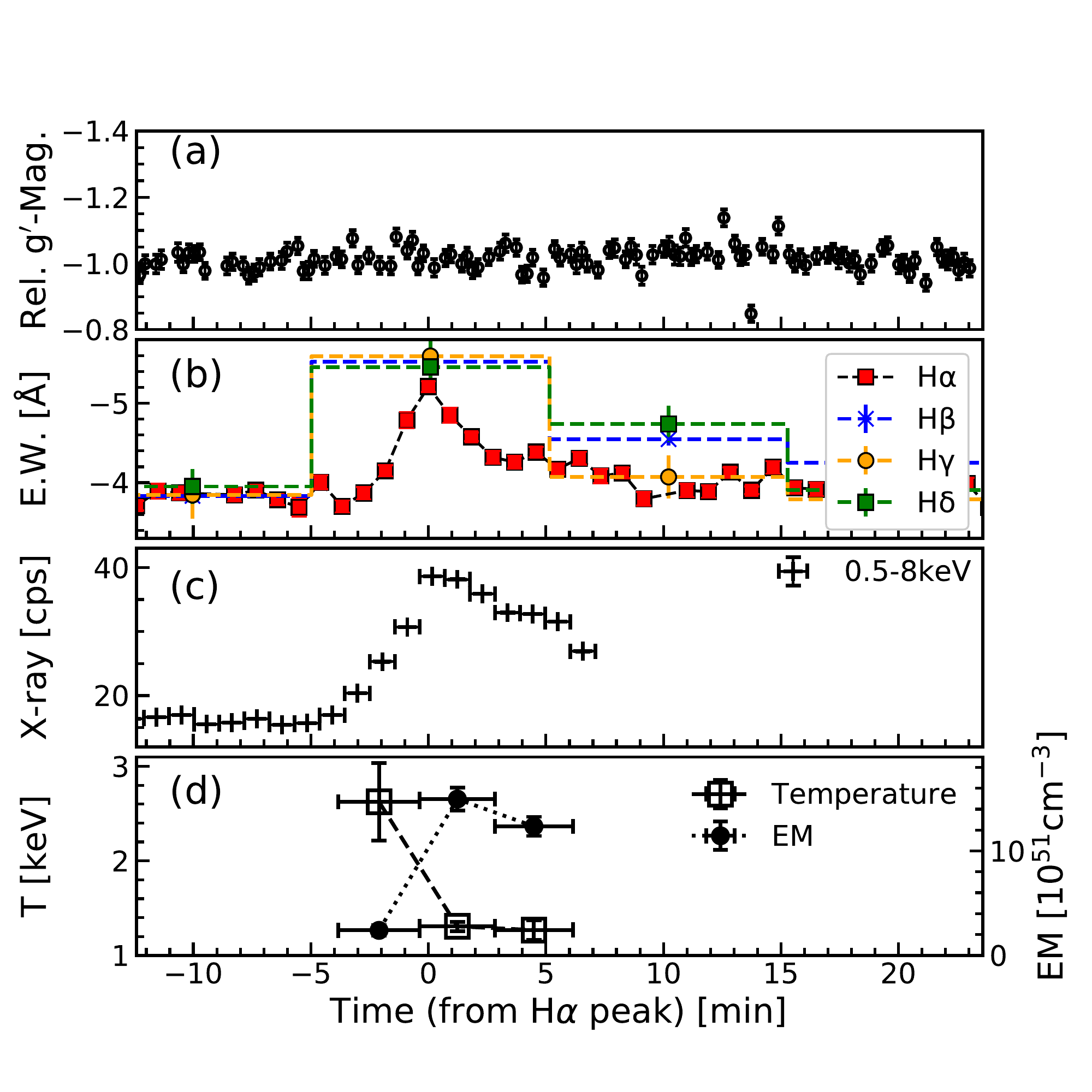}
\end{center}
\caption{ Light curves of the flare \#4 observed in $g$-band (MITSuME, panel a), Balmer lines (Seimei Telescope and SCAT, panel b), and X-ray (NICER, panel c). The panel (d) shows time variations of the emission measure and temperature during the flare. The values are derived from the pre-flare subtracted X-ray spectra (0.5-8.0 keV).  }
\label{fig:sf4}
\end{figure}

Figure \ref{fig:sf4} shows the light curve of flare \#4, and there are observations by all instruments (Seimei/KOOLS-IFU, SCAT, Nayuta/MALLS, MITSuME, and NICER) during flares.
Enhancement of Balmer lines and X-ray are clearly detected, but the continuum emissions are too weak to detect with the MITSuME photometric sensitivity.\footnote[6]{As we described, the MITSuME CCDs have the inevitable noise pattern, and the flat flaming can make the photometric sensitivity worse.}
The enhancement of the equivalent width is the same for all Balmer lines, which is different from the flare \#2 although the enhancement level is similar to each other.
The total radiated energy in H$\alpha$, H$\beta$, H$\gamma$, and H$\delta$ are calculated to be 1.1$\times10^{30}$ erg, 1.6$\times10^{30}$ erg, 5.0$\times10^{29}$ erg, and 9.6$\times10^{29}$ erg, respectively.

For the X-ray data, the rise, peak, and initial decay phase were successfully observed.
The count rates become about twice the quiescent values.
The results of the model fitting of the flare spectra are also plotted in the panel (d) of Figure \ref{fig:xspec4}.
The emission measure and temperature at the flare peak are derived as 1.15$\times10^{52}$ cm$^{-3}$ and 1.57 keV  (18.2 MK),  respectively.
After the temperature  increases initially,  the emission measure increases later, which is similar to the typical X-ray behavior accompanied with chromospheric evaporations in solar flares (e.g., \cite{2002ApJ...577..422S}).  
The observed X-ray flare energy in  0.5 -10 keV  is calculated to be 3.4$\times10^{31}$ erg, which is larger by about one order of magnitude than the Balmer line energy.
 Note that even in the initial phase, no significant hard X-ray power-law component is detected and the spectra can be fitted only with the single component thermal spectra. 

\section{Rotational Modulation} \label{sec:4}

\begin{figure}[htbp]
\begin{center}
\includegraphics[scale=0.4]{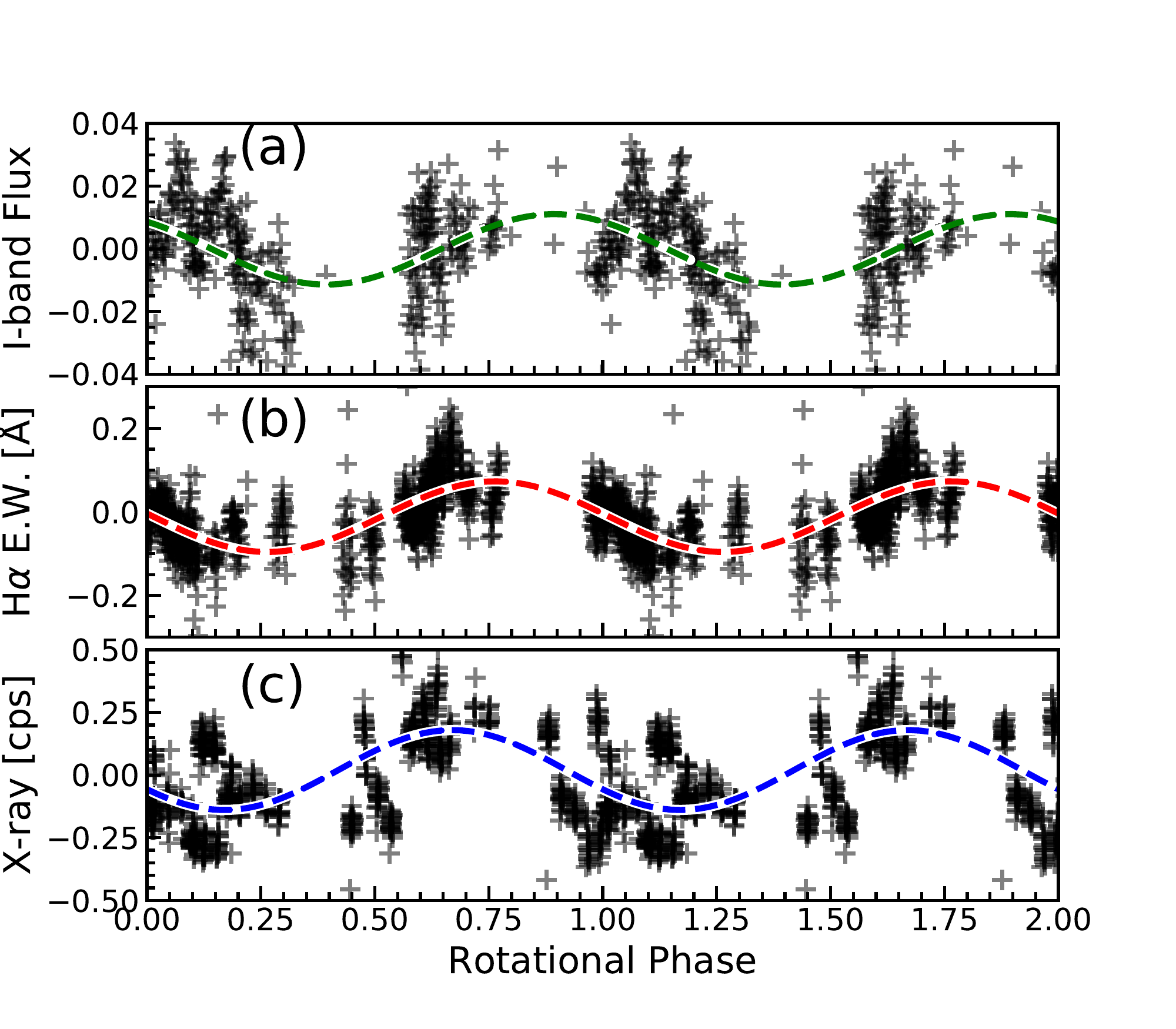}
\end{center}
\caption{Light curves  relative to the mean value   of (a) I-band continuum, (b) H$\alpha$, and (c) soft X-ray folded by rotational period $\sim$ 2.24 days.  Note that flares were removed from the panel.  Two rotational phases are plotted. The dotted line in each panel indicates the best-fit sinusoidal curve.}
\label{fig:rotmoduration}
\end{figure}

Figure \ref{fig:rotmoduration} (a-c) show the light curves which are folded with the rotational period of 2.2399 days reported in the previous work  of \citet{2012PASP..124..545H}. 
We removed the visibly-checked flares to make Figure \ref{fig:rotmoduration}.
The H$\alpha$ and X-ray phase-folded light curves show the clear periodic feature, which would be the signature of the rotational modulations of the AD Leo with the bright active region in chromosphere and corona.
On the other hand, the photometric light curves do not show clear periodicity probably due to the lack of photometric sensitivity.
The phase-folded light curves are fitted with a simple sinusoidal curve indicated with the dotted lines.
The phase difference between X-ray and H$\alpha$ is only 0.094, and that between X-ray and I-band continuum is 0.22.
Although the I-band periodicity is not clear, the phase of X-ray and H$\alpha$ periodicity seems to be highly correlated.

\begin{figure}[htbp]
\begin{center}
\includegraphics[scale=0.4]{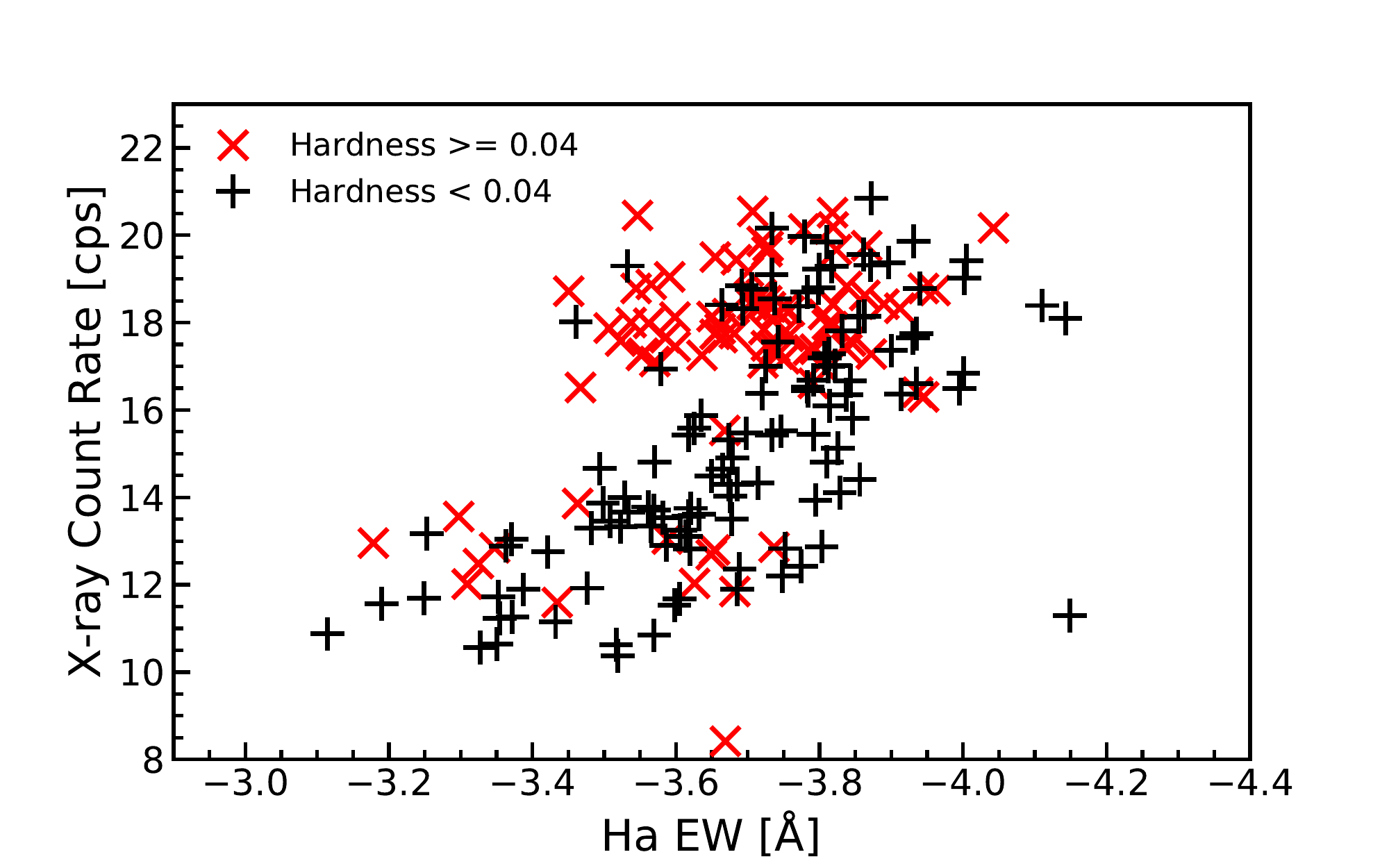}
\end{center}
\caption{Comparison between H$\alpha$ E.W. and X-ray count rates (0.5-8keV) during non-flaring phase. The red pluses and black crosses show the values where the hardness ratio values (i.e. F(2-8 keV)/F(0.5-2 keV)) are $<$ and $>$ 0.04.}
\label{fig:rotmoduration2}
\end{figure}

Figure \ref{fig:rotmoduration2} shows the comparison between X-ray and H$\alpha$ intensity where both of the data exist at the same time.
The X-ray and H$\alpha$ intensity have a positive correlation, while there looks to be no correlation between them for the high hardness ratio.
The amplitude of the X-ray modulations (16 \%) is twice larger than that of the H$\alpha$ modulation (8 \%), which may be caused by the filling factor or contrast of the active regions.

\section{Radiative-hydrodynamic Flare Modeling}\label{sec:5}

In Section \ref{sec:3}, we found that, (1) accompanied by the large white-light enhancement, the H$\alpha$ emission line width dramatically increases to 14 {\AA} from 8 {\AA}, and that (2) some weak H$\alpha$ flares are not accompanied with white-light emission.
These observational features motivate us to carry out numerical modelings of stellar flares to know what happens in the atmosphere.
In this following section, we report the result of one-dimensional radiative-hydrodynamic (RHD) flare modelings with the RADYN code (\cite{1992ApJ...397L..59C}, \yearcite{1995ApJ...440L..29C}, \yearcite{1997ApJ...481..500C}, \yearcite{2002ApJ...572..626C}), which calculates
hydrogen, helium, and Ca II in non-LTE framework and with non-equilibrium ionization/excitation.
We refer the reader to \citet{2015ApJ...809..104A}, \citet{2015SoPh..290.3487K}, and \citet{2017ApJ...837..125K} for extensive descriptions of the flare simulations. 

\subsection{RADYN Flare Model Setup} \label{sec:5.1}

The pre-flare atmosphere  (log$g$ = 4.75)  for an M dwarf in our modeling is described in Appendix A of \citet{2017ApJ...837..125K}.
The pre-flare coronal electron density becomes up to $10^{11}$ cm$^{-3}$. 
Although this value is larger by one or two orders of magnitude than that of solar atmosphere, it is approximately consistent with the stellar X-ray observations (e.g., \cite{2006ApJ...647.1349O}).

Several improvements have been made to the RADYN flare code since \citet{2015ApJ...809..104A}, which are worth noting here (they will be described further in \cite{A20}, in prep). The hydrogen broadening from \citet{2017ApJ...837..125K} and \citet{2009ApJ...696.1755T} have been included in the dynamic simulations; since we are comparing to  H$\alpha$  observations, this update to the hydrogen broadening is a critical improvement (\cite{K20}, in prep).  The pre-flare atmosphere was relaxed with this new hydrogen broadening, and we choose to use the X-ray backheating formulation from \citet{2005ApJ...630..573A} for these models (\cite{K20}, in prep); the resulting pre-flare apex temperature is $\sim$ 3 MK, with electron density $\sim$ 10$^{11}$ cm$^{-3}$. Finally, we used a new version of the F-P solver (\cite{A20}, in prep), which gives a moderately smoother electron beam energy deposition profile over height in the upper chromosphere. These changes have been implemented for the solar flare models presented in \citet{2020arXiv200405075G}.

Recently, this kind of RHD simulations has been widely carried out for the modeling of solar flares and dMe flares.
However, the H$\alpha$ behaviors have been not so much well-investigated partly because it is difficult to understand due to its NLTE formation properties and large opacity variations over the line profile. 
In this study, we revisit the basic properties of the H$\alpha$ lines for dMe flares, and our aim of the numerical simulation is (1) to understand the H$\alpha$ line width behaviors as a function of the injected energy, and (2) to understand the basic relation between H$\alpha$ and optical continuum.

\subsection{Flare Heating Inputs} \label{sec:5.2}
 In this section, we introduce the flare heating parameters used in our simulations. 
We aim to know the H$\alpha$/continuum intensity and H$\alpha$ line width as a response to the flare heatings. 
In this study, we performed two kinds of simulations in which lower atmospheres are heated by (i) the non-thermal high energy electrons (e.g., \cite{2017ApJ...837..125K}) and (ii) thermal conduction from the heated loop-top  (e.g. \cite{1997ApJ...489..426H}; \cite{1989ApJ...346.1019F}; \cite{2017ApJ...837..125K}). 
We performed the non-thermal and thermal simulation separately to see their difference in the behavior of the emission atmosphere.
Note that we consider only one-dimensional tube in this study, but non-thermal line broadening among multi-loops (e.g., \cite{2011A&A...534A.133F}) may have to be also considered in future  (e.g., \cite{2006ApJ...637..522W}; \cite{2017ApJ...837..125K}). 

In the case of non-thermal heating, the energy flux density of non-thermal electrons with a power-law profile ($F(E) \propto E^{-\delta}$) is injected from the loop top. 
The important parameters are total energy flux density ($F_{\rm NT}$), the lower energy cutoff ($E_{\rm C}$), and the power-law index ($\delta$).
Again, we aim to know the H$\alpha$ /continuum intensity and line width by controlling these parameters.
For simplicity, the $E_{\rm C}$ is chosen to be 37 keV \citep{2015ApJ...809..104A}.
$F_{\rm NT}$ is chosen to be 10$^{10}$ cm$^{-2}$ s$^{-1}$ (F10), 10$^{11}$ cm$^{-2}$ s$^{-1}$ (F11), 10$^{12}$ erg  cm$^{-2}$ s$^{-1}$ (F12) (hereafter, we express the flare model with $F_{\rm NT}$ of 10$^{10}$ erg  cm$^{-2}$ s$^{-1}$ as F10),  following the values simulated in previous works of large solar flares (\cite{2003ApJ...595L..97H}, \cite{2005ApJ...630..573A}, \yearcite{2006ApJ...644..484A}, \cite{2015SoPh..290.3487K}),  and the power-law index $\delta$ is determined to be 3 and 5.
The heating profile is assumed to be a triangle time variation with the peak time of 2 s where energy flux linearly increases and decreases in the rising and decay phase respectively.
%The set up is summarized in the Table.

In the cause of thermal-conduction heating, the heating source term is input to the energy equation at the magnetic loop top (e.g. \cite{1997ApJ...489..426H}).
The energy flux of the thermal source is 5$\times$10$^{10}$ cm$^{-2}$ s$^{-1}$ (5F10), 10$^{11}$ cm$^{-2}$ s$^{-1}$ (F11), 5$\times$10$^{11}$ cm$^{-2}$ s$^{-1}$(5F11), 10$^{12}$ erg  cm$^{-2}$ s$^{-1}$ (F12) per each magnetic loop.
The heating profile is also a triangle time variation with the peak time of 4 s which is twice the non-thermal case.

\subsection{Simulation Result I: H$\alpha$ Line Broadening for Non-thermal/Thermal Heating} \label{sec:5.3}

\begin{figure*}[htbp]
\begin{center}
\includegraphics[scale=0.35]{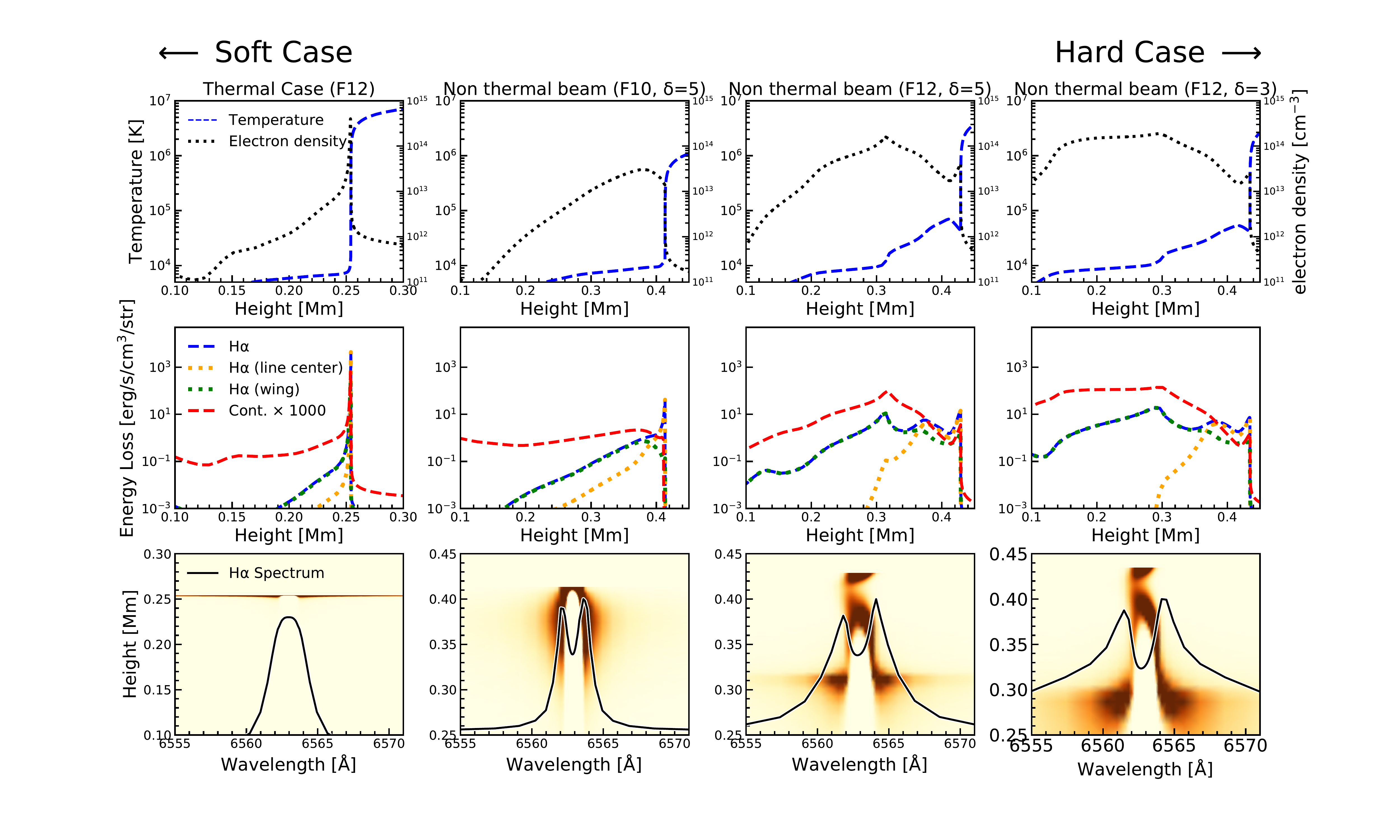}
\end{center}
\caption{ Examples of the flaring atmospheric condition at the flare peak calculated by the RADYN code for M-type stars. From left to right, soft to hard energy input cases are shown. Top: the atmospheric temperature (dashed blue line) and electron density (dotted black line) as a function of height from the photosphere at the flare peak of the H$\alpha$ radiation. Middle: contribution functions for H$\alpha$ and 6690 {\AA} continuum as a function of height. For the H$\alpha$, the contribution from line core and wing are separately plotted. Bottom: background show the normalized H$\alpha$ line contribution function in the space of wavelength and atmospheric height. The black curves show the H$\alpha$ spectra. }
\label{fig:radynatmosphere}
\end{figure*}

 In this section, we show the H$\alpha$ line broadening for the non-thermal/thermal heatings as a response to the different kine of simulation input parameters.
First, we show where the H$\alpha$ line and continuum emissions come from in the one-dimensional flare atmosphere  in Figure \ref{fig:radynatmosphere}.
Four typical results of the simulated flare atmospheres are shown in Figure \ref{fig:radynatmosphere}.
The upper panels show the detailed atmospheric structure, the middle panels show the contribution function of H$\alpha$ and continuum, and bottom panels show the line formation region of the H$\alpha$. 
In the middle panels, you can see that the H$\alpha$ (wing and center) and optical continuum enhancements come from the upper to lower chromosphere in the non-thermal electron case (hard case; the right side of Figure \ref{fig:radynatmosphere}), while they mainly come from the upper chromosphere and transition region in the thermal case  (soft case; the left side of Figure \ref{fig:radynatmosphere}).

\begin{figure}[htbp]
\begin{center}
\includegraphics[scale=0.4]{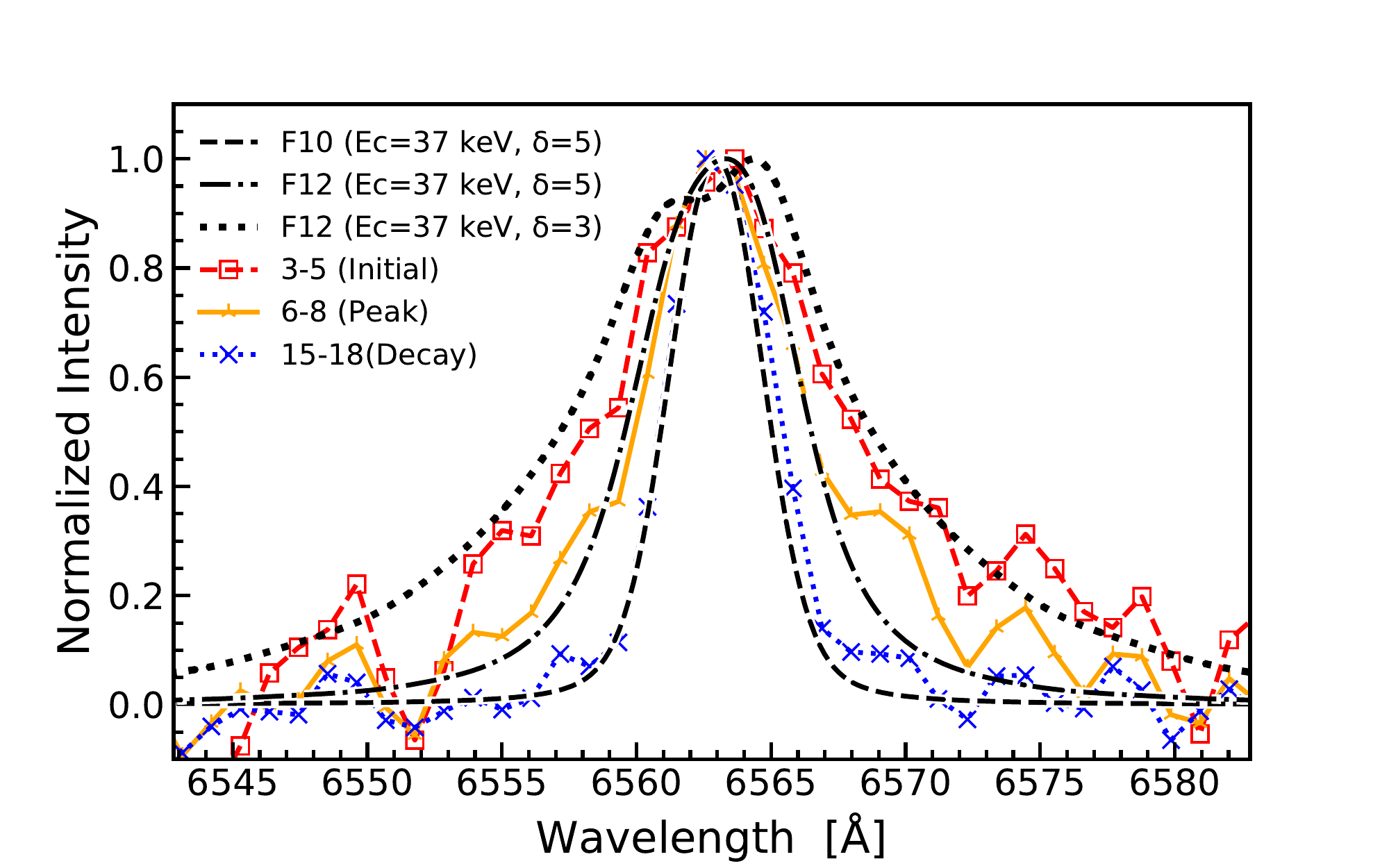}
\end{center}
\caption{Comparison of H$\alpha$ spectra between observations (colored) and model (black). For the observational spectra, the temporal evolution of H$\alpha$ spectra of flare \#1  (panel (B) of Figure \ref{fig:flare1spectra})  are plotted. For the model spectra, F10 ($\delta$=5), F12 ($\delta$=5), and F12 ($\delta$=3) cases are plotted as references. The model spectra are convolved with the Gaussian-function instrumental profile of Seimei/KOOLS-IFU (R$\sim$2000, $\Delta \lambda \sim$ 3 {\AA}; \cite{2019PASJ...71..102M}),  and the central reversal seen in Figure \ref{fig:radynatmosphere} is not resolved in the low-resolution spectra. }
\label{fig:B}
\end{figure}

 Next, let us simply compare the simulated spectra with the observations obtained in the Section \ref{sec:3}.
Figure \ref{fig:B} shows the comparison of the H$\alpha$ spectra between observations and models.
For the initial phase (red), the observed line shape is more consistent with the hard- and high-energy spectrum model of F12 ($\delta$ = 3).
For the later phases, the observed line shape is more consistent with the weaker (F10; $\delta$ = 3) or softer (F12; $\delta$ = 5) energy spectrum models.
Of course, all the observed spectra is not corresponding ``flare peak" (including rising/decay phases) and the readers may think that it is better to compare them with time-dependent H$\alpha$ spectral evolution for a given simulation in a single loop. 
However, in analogy with solar flares, the stellar flares are expected to be observed as a superposition of many magnetic loops and each footpoint has a different ``flare peak".
Therefore, under the assumption that one loop quickly decays, the comparison with the peak spectra of each model setup is not so much bad, and it is easy to derive the physical parameters of the energy injection.
As a result of comparison between observations and models, one can say that in the initial phase, the high energy electron with large energy deposition rate and hard spectral distribution occurred, and the spectral feature changes to softer/weaker energy injection.

\begin{figure}[htbp]
\begin{center}
\includegraphics[scale=0.4]{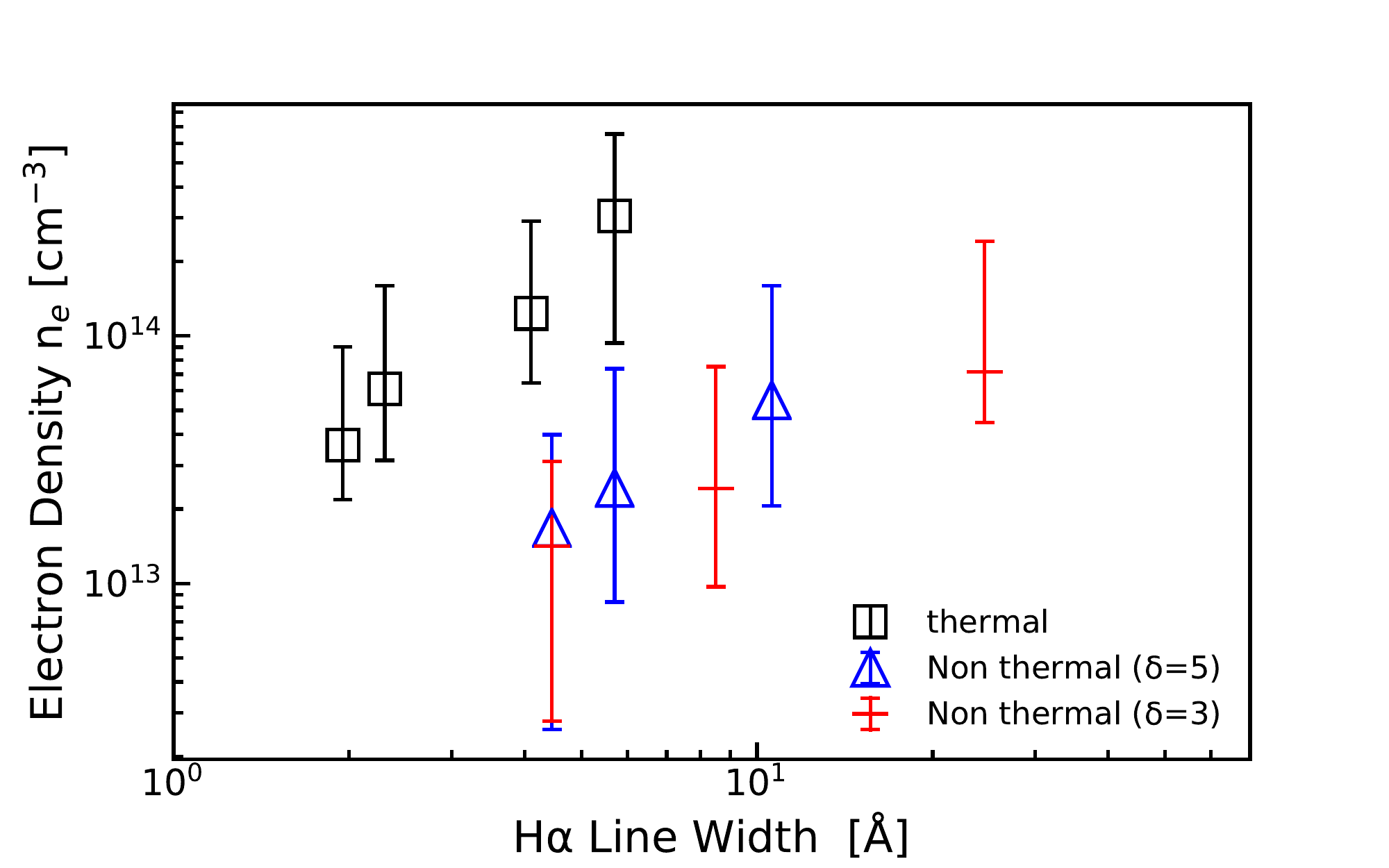}
\end{center}
\caption{Comparison between the H$\alpha$ intensity and electron density weighted by the contribution function of the H$\alpha$  obtained by the RADYN simulations.  The squares show the thermal case. The crosses and triangles show the non-thermal electron cases, and the different symbols indicate the different power-law index. As input energy flux (F{\#}; from F10 to F12) increases, the line width and electron density increase. }
\label{fig:A}
\end{figure}

 Finally, we show the relation between H$\alpha$ line broadening and some physical parameters such as atmospheric density or energy injections.
Figure \ref{fig:A} shows the comparison between H$\alpha$ line width and electron density in the chromospheric condensation region at the flare emission peak for each case.
For the thermal cases, the emission of H$\alpha$ is mostly radiated from the upper chromosphere and transition region where there is less self-absorption\footnote[9]{ Here, self-absorption means that the emissions are absorbed by the upper atmosphere having large opacity not to be able to go out to the surface. The upper layer of the chromosphere/transition region (i.e. corona) have very small opacity in continuum and Balmer lines.  }, so the positive relation for the thermal cases is likely to come from the linear Stark effect \citep{2017ApJ...837..125K}.
For the non-thermal cases, the harder spectral cases show wider line broadening.
If the model atmosphere in F12 ($\delta=3$) is compared with that in F12 ($\delta=5$) in Figure \ref{fig:radynatmosphere}, only the line wing contributions are enhanced in the deep chromosphere for  the harder beam of F12 ($\delta$=3). 
This would be because the hard high energy electrons deposited in the deep chromosphere causes the strong Stark effect and self absorptions.
 In cases where the electron density is not so much  different for each case, the self-absorption can largely contribute to these differences.

\subsection{Simulation Result II: Relation between Balmer-lines and Optical-continuum Emissions  -- What Are Non-white-light flares Like? } \label{sec:5.4}

\begin{figure}[htbp]
\begin{center}
\includegraphics[scale=0.4]{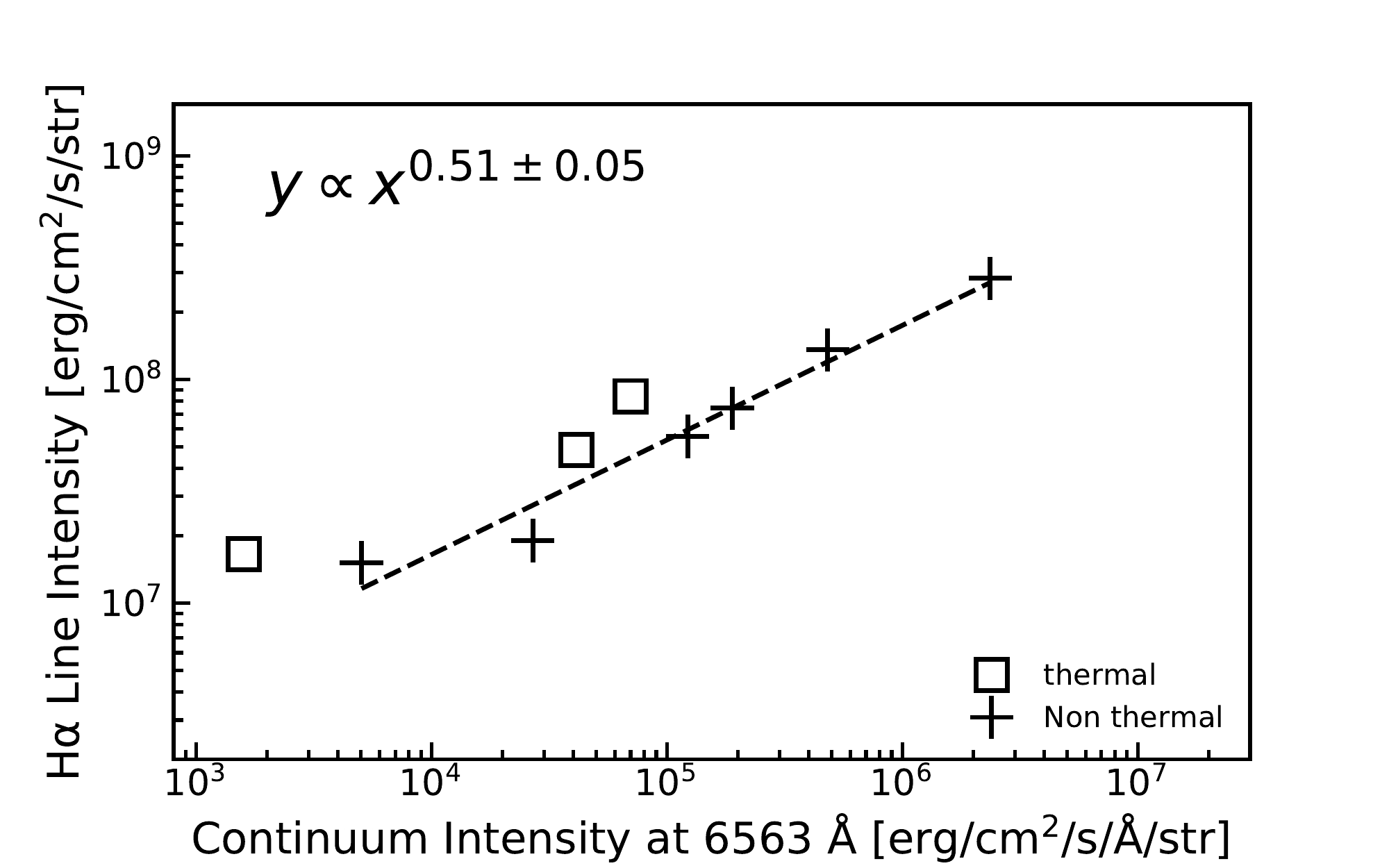}
\end{center}
\caption{Comparison between the H$\alpha$ intensity and continuum intensity at 6563 {\AA}  obtained by the RADYN simulations.   The H$\alpha$ intensity is  wavelength-integrated across the line.  Square symbols show the thermal case, and cross symbols show the non-thermal electron cases.  As input energy flux (F{\#}; from F10 to F12) increases, H$\alpha$ and continuum intensities increase.  The dotted line is  the fitted one for the non-thermal case, and the power-law index is derived to be 0.51. Note that the 5F10 thermal case show very week continuum emission, and is not plotted in this figure.}
\label{fig:hacont}
\end{figure}

 In this section, we show the relation between H$\alpha$ lines and optical continuum emissions as a response to the different kinds of input parameters of flare simulations. 
Figure \ref{fig:hacont} shows the comparisons between the continuum and H$\alpha$ emission in the RADYN simulations.
We found that the relation between the optical continuum and H$\alpha$ emission is not linear, but expressed as $I_{\rm H\alpha} \propto I_{\rm cont}^{\alpha}$, where  $\alpha = 0.51 \pm 0.05$. 
We consider that there are the following two possibilities for this nonlinear relation: (1) an opacity difference between optically-thick H$\alpha$ and optically thin continuum in the chromosphere, and (2) an emissivity difference between H$\alpha$ and continuum   (e.g., \cite{2019ApJ...878..135K}), (3) or both.   

\begin{figure}[htbp]
\begin{center}
\includegraphics[scale=0.4]{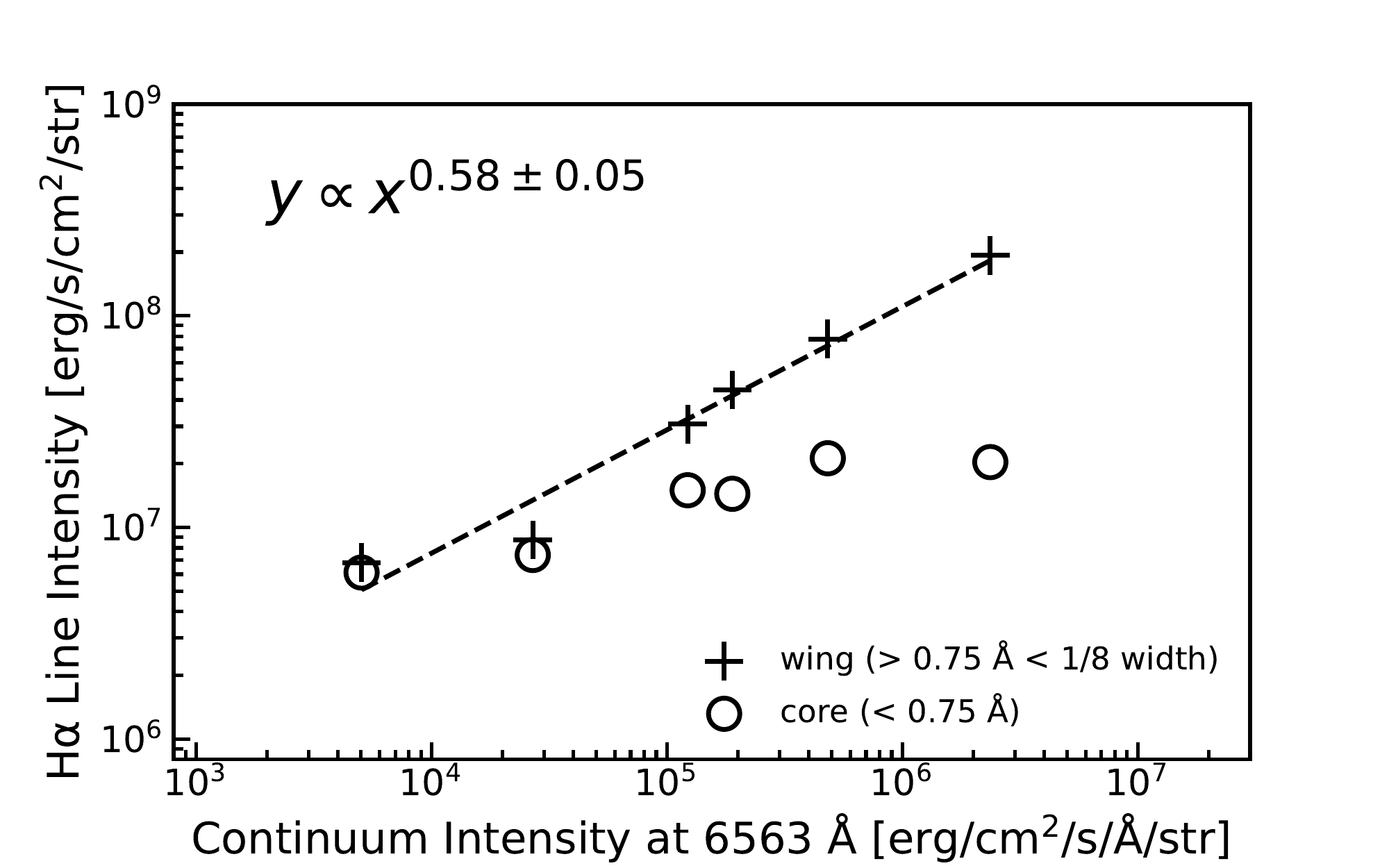}
\end{center}
\caption{Comparison between the H$\alpha$ intensity and continuum intensity at 6563 {\AA} for the non-thermal case  obtained by the RADYN simulations.  The contribution from line core (circles; $< 0.75 {\AA}$ from the line center) and wing (crosses; emission integrated over 0.75 {\AA} from line center to 1/8 width) are separately plotted here. The dotted line is the fitted one for line wing, and the power-law index is derived to be 0.58.  As input energy flux (F{\#}; from F10 to F12) increases, H$\alpha$ and continuum intensities increase.  Since the line center is optically thick compared to the line wing, the emergent intensity of line center cannot become large when the input energy becomes large compared to line wing. }
\label{fig:hacontsep} 
\end{figure}

In the case of the possibility (1), since H$\alpha$ is optically thick in the chromosphere, the line shape heavily suffers from the self-absorption,  especially  in line center. 
Therefore, the more energetic the flare input is, the less H$\alpha$ emission ($\eta_{H\alpha}$) comes from the lower flaring atmosphere compared with the optical continuum which is optically thin  and can escape from a large range of heights. 
In Figure \ref{fig:hacontsep}, the comparisons between the continuum and H$\alpha$ emission are plotted for H$\alpha$ line wing and center separately.
You can see that the H$\alpha$ line center is less sensitive to the flare input and continuum emission.
This is because the line center is more optically thick than the line wing, and self-absorption more or less affects the nonlinearity  (for the absorption line, see e.g. Figure 9.1 in \cite{2003rtsa.book.....R}). 

\begin{figure}[htbp]
\begin{center}
\includegraphics[scale=0.4]{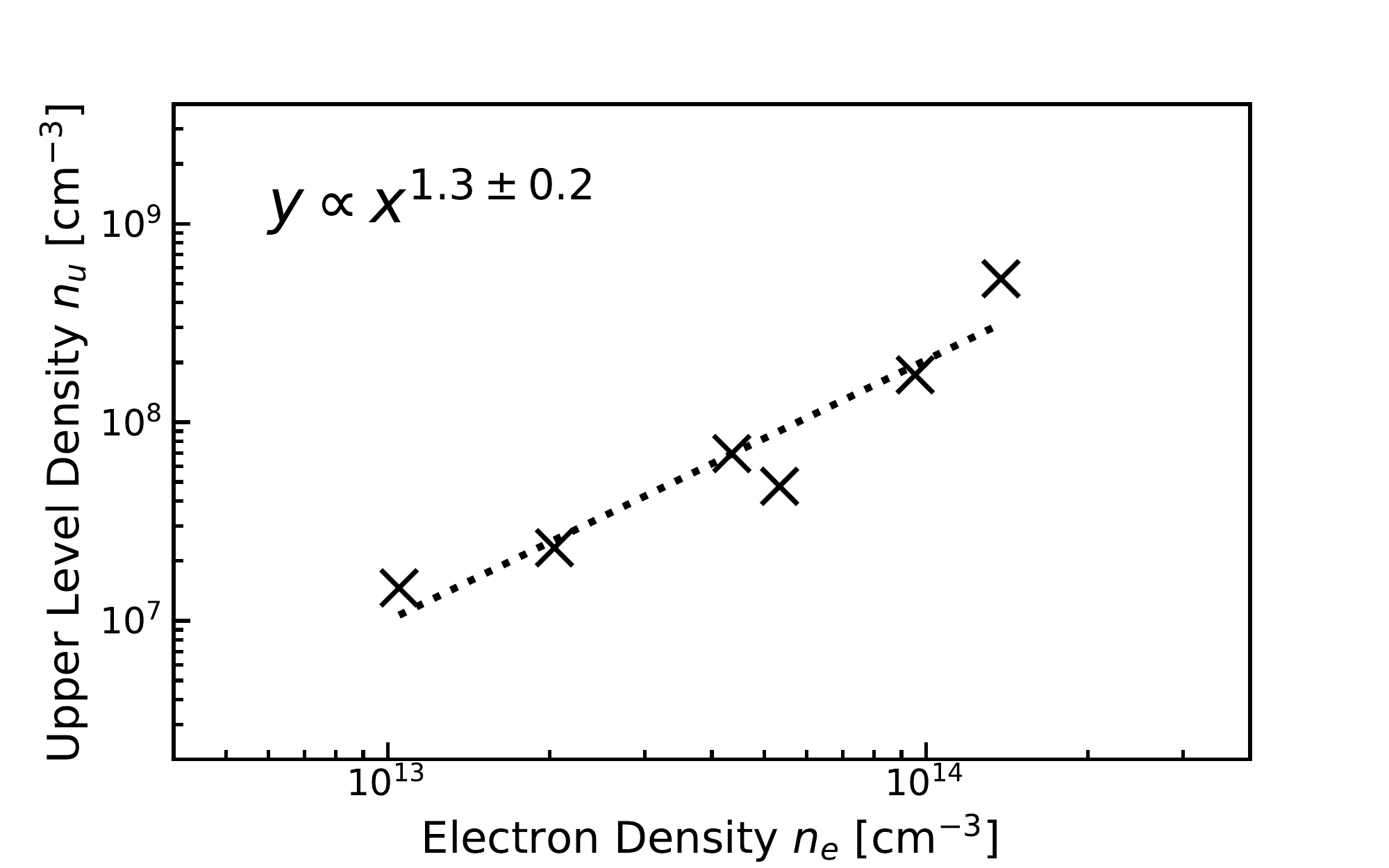}
\end{center}
\caption{Comparison between H$\alpha$ upper level density and electron density weighted the line wing contribution function  obtained by the RADYN simulations.  The dotted line is the fitted one, and the power-law index is derived to be 1.3.}
\label{fig:nenu}
\end{figure}

In the case of the possibility (2), the emissivity of the H$\alpha$ can be expressed as
\begin{equation} \label{eq:1}
\eta_{H\alpha} = \frac{h\nu_{\rm 0}}{4 \pi}n_{\rm u}A_{\rm u}\phi (\nu -\nu_{\rm 0} )  \propto n_{\rm u}
\end{equation}
where  $\eta_{H\alpha}$ is the emissivity of H$\alpha$, $n_{\rm u}$ is the upper level Hydrogen density (n=3), $h$ is a Planck constant, $\nu_{\rm 0}$ is the line center frequency, $A_{\rm u}$ is an Einstein constant, and $\phi$ is the profile function for H$\alpha$.
On the other hand, the emissivity of optical continuum ($\eta_{\rm cont}$) can be expressed as
\begin{eqnarray} \label{eq:2}
\eta_{\rm cont.} = n_{\rm e}n_{\rm p}F(T, \nu) \propto n_{\rm e}^2
\end{eqnarray}
where  $n_{\rm e}$ is the electron density, $n_{\rm p}$ is the proton density, and $F(T, \nu)$ is the function of temperature $T$ and frequency $\nu$  (for the detailed expression, see \cite{2019ApJ...878..135K}), if we assume the Paschen continuum is dominant in the optical range.
If we assume that the electron density is roughly proportional to the upper-level density of H$\alpha$ and temperature is not so much different in the chromospheric layers, we can deduce $\eta_{\rm H\alpha} \propto \eta_{\rm cont}^{\alpha}$, where $\alpha \sim 0.5$, which may be able to explain the non-linearity.
Figure \ref{fig:nenu} shows the comparison between $n_e$ and $n_u$ which are weighted by the continuum and line-wing contribution function, respectively.
Considering the obtained relation  $n_u \propto n_e^{1.3 \pm 0.2}$, the equations (\ref{eq:1}) and (\ref{eq:2}) reduce to $I_{\rm H\alpha} \propto I_{\rm cont}^{\alpha}$, where $\alpha \sim 0.65$  (with $\tau_{\rm wing}$ $\ll$ 1 in chromosphere),  which is simlar to the relations for line wing $I_{\rm H\alpha wing} \propto I_{\rm cont}^{\alpha}$, where  $\alpha = 0.58 \pm 0.05$.

These results mean that the non-linear relation between H$\alpha$ and continuum intensity comes from both of the (1) opacity effect and (2) emissivity difference.
This non-linearity means that  the emergent continuum intensity more significantly decreases than the emergent H$\alpha$ intensity  when the energy input rate into the chromosphere decreases.
 For example, the H$\alpha$ intensity decreases by a factor 3 whereas the continuum intensity decreases by one order of magnitude. 
This can explain why there is no white-light emission on the relatively weak H$\alpha$ flares:  white-light emission is difficult to detect compared to the H$\alpha$ emission in the case of weak flares. 
%Although there would be other possibilities for the explanations on the non-white-light flares as discussed in the next section, the 1-d RADYN simulations propose another possibility that energy partitions can be different between H$\alpha$ and optical continuum.
%In fact, \ref{2017ApJ...850..204W} 

\section{Discussion and Conclusion} \label{sec:6}

\subsection{Stellar Flares  -- Observations and Simulations} \label{sec:6.1}
Let us summarize the observational results in this work.
We performed monitoring observations on an M-dwarf flare star AD Leo during 8.5 night with the Seimei Telescope, SCAT, NICER, and the collaborations of OISTER program.
Twelve flares were detected and multi-wavelength data were obtained for several flares (see Section \ref{sec:3} and Appendix \ref{app:1}). 
Particularly, flare \#1 was a superflare whose energy is 1.4$\times10^{33}$ erg in the $g'$-band filter (the total radiated energy in all the optical filters was 1.9$\times10^{33}$ erg).
We found the following three interesting events:
\begin{enumerate}
\item during the superflare (flare {\#1}), the H$\alpha$ emission line full width at 1/8 maximum dramatically increases to 14 {\AA} from 8 {\AA} accompanied with the large white-light flares (Section \ref{sec:3.2}),
\item some weak H$\alpha$ and X-ray flares (e.g. flare {\#4}) are not accompanied with white-light emissions which are candidates  of so-called non-white-light flares in the case of solar flares (Section \ref{sec:3.4}), 
\item clear rotational modulations are found in X-ray/H$\alpha$ in the same phase whereas the continuum periodicity is not clear due to the photometric sensitivity (Section \ref{sec:4}, discussed in Section \ref{sec:6.2}).
\end{enumerate}
We performed a one-dimensional RHD simulation with RADYN code to understand the behavior of H$\alpha$ and optical continuum obtained in the above points (1) and (2) (Section \ref{sec:5}).

%%Ha line width

As for the point (1), the numerical simulation (Section \ref{sec:5.3}) shows that the line width of H$\alpha$ largely depends on both the energy flux density ($F$) and the energy spectrum (a power-law index $\delta$).
By changing the energy spectra, the degree of contribution from Stark broadening and opacity broadening does change, meaning that it is difficult to constrain the input energy spectra from only the H$\alpha$ spectra of the superflare {\#}1.
Even if the additional information such as the continuum fluxes is given, there is another degeneracy between intensity ($I$) and emission area ($A_{\rm flare}$), which makes it difficult to constrain the energy spectra.

Considering the large H$\alpha$ broadening during the superflare, it would be at least possible to say that the H$\alpha$ broadening in the initial phase of the superflare indicates a hard- and/or high-energy flare injection via non-thermal electron like the case with $F_{\rm NT}$ = 10$^{12}$ erg  cm$^{-2}$ s$^{-1}$ and $\delta$ = 3, and the decrease in the line width indicates the decrease of the energy flux density and/or the softening of energy spectra  at different locations in the flare  ribbon. 
Previous studies also indicate that  $F_{\rm NT}$ = 10$^{12}$  erg  cm$^{-2}$ (low-energy cutoff $\gg$ 37 keV) or 10$^{13}$ erg  cm$^{-2}$ s$^{-1}$  is necessary to reproduce the broad-band continua of M-dwarf flares \citep{2016ApJ...820...95K}, and fluxes larger than this  are sometimes inferred  even on the Sun \citep{2011ApJ...739...96K}.
More comparison between observations and modelings can give us a clue to the universality or difference of the particle acceleration on solar and stellar flares.

%%NWFL??
As for the point (2), our simulation also shows that the H$\alpha$ and optical continuum intensity have non-linear relation $I_{\rm H\alpha} \propto I_{\rm cont}^{\alpha}$, where $\alpha \sim 0.5$, which is largely caused by the opacity and emissivity difference (Section \ref{sec:5.4}). 
This non-linearity can contribute to the cause of non-white-light flares: as the energy input rates decrease, the continuum emissions more significantly decrease than H$\alpha$ emissions.
In the case of the solar flare, it is reported that non-white-light flares tend to have long durations, i.e. small energy deposition rates  (\cite{1987ApJ...322..999C}; \cite{2003A&A...409.1107M}; \cite{2017ApJ...850..204W}),  which is consistent with our simulation  and interpretation/analysis. 
However, it would not be consistent with the relatively a short duration $\sim$ 7 minutes of flare {\#4}, and we need more samples of stellar non-white-light events (e.g. Maehara et al. in preparation) and the validation in spatially-resolved solar flares would be required.
As another possibility, stellar non-white-light flares may be explained by a flare over the limb.
In this case, if we assume that white-light emissions originate from only chromosphere/photosphere like solar flares, the white-light emission source at loop footpoints is invisible while the X-ray and H$\alpha$ emission are visible in the flare loop in the corona.
 However, this possibility may be less likely because white-light emissions can be visibly emitted from the dense (post) flare loops (\cite{2018ApJ...859..143H}; \cite{2018ApJ...867..134J}) even though the footpoints are invisible over the stellar limb.

%%Rotational Modulation
\subsection{Rotational Modulations} \label{sec:6.2}
As summarized in the point (3) of Section \ref{sec:6.1}, we found X-ray and H$\alpha$ rotational modulations on AD Leo with a period of 2.24 days in almost the same phase (see Section \ref{sec:4}).
It is to our knowledge rare to simultaneously detect the rotational modulation in coronal and chromospheric emissions on M-dwarf, although some previous studies are showing rotational modulations with broadband optical and X-ray \citep{2017MNRAS.464.3281W} or with only X-ray (e.g. \cite{2003A&A...407L..63M}; \cite{2007MNRAS.377.1488H}).
The correlation between chromospheric and coronal emission would be because both H$\alpha$ and X-ray would come from the magnetically-active regions.
Although the small difference of rotational phase ($\sim$ 0.1) between H$\alpha$ and X-ray may indicate the difference of the visibility or location of the bright active region in each wavelength, it may be due to the sparse data sampling and/or the very rough fitting by a sine curve. 
The amplitude of the brightness variation in X-ray and H$\alpha$ is 32 {\%} and 16 {\%}, respectively. 
%However, if the I-band amplitude of 6{\%} is true, it looks rather large considering the low inclination angle of AD Leo (Morin, 2008). 
The factor-of-two difference in amplitude between X-ray and H$\alpha$ may be due to that in active region filling factor between chromosphere and corona, but may be due to the contrast between the active region and quiet regions.
\citet{1985ApJ...299L..47S} however inferred that the 73 {\%} of the surface of AD Leo is covered by active regions outside of dark spots.
% containing a mean field strength of 3800 G, which would be bright in chromospheric lines and possibly in X-ray, and large brightness variation is therefore expected.
One possibility to explain the difference between the observed brightness amplitude in X-ray/H$\alpha$ and the reported filling factor is that the low inclination angle of AD Leo ($i$ = 20$^{\circ}$; \cite{2008MNRAS.390..567M}) reduce the brightness amplitude, and the other is that the large filling factor significantly reduced the stellar brightness amplitude (\cite{1994ApJ...420..373E}; \cite{2020ApJ...890..121S}). 

Although the rotational brightness variations were detected for X-ray, we could not detect clear rotational variations in the hardness ratio of X-ray band which is related to the coronal temperature: the spectra show the relatively low hardness ratio even in the high X-ray intensity.
This could be because the temperature is not sensitive to the magnetic loop size (\cite{1978ApJ...220..643R}), or may be because the active region consists of group of small magnetic loops and therefore the increase in the number of magnetic loops does not affect the changes in coronal temperature. 

The optical rotational modulation is less than $\sim$6 {\%} possibly due to the lack of sensitivity, but maybe it is true considering that optical variability of M dwarfs like GJ 1243 (dMe4.0V) is comparable with this amplitude in Kepler (e.g. \cite{2014ApJ...797..121H}; \cite{2014ApJ...797..122D}). 
However, considering this low inclination angle, the I-band amplitude of 6{\%} looks rather large, and it is interesting to try a photometric spot modeling on the data (e.g., \cite{2020ApJ...891..103N}). 
The optical rotational phase is not completely anti-correlated with X-ray/H$\alpha$ ones (0.2--0.3 rotational phase), which may indicate the spotted side of the hemisphere is not the same as coronal/chromospheric active regions, although more precise measurements of the optical rotational modulation would be necessary for conclusions.  
In this observational period, it is difficult to estimate the active region emergence or decay, but the long-term continuous monitoring observation with multi-wavelength can reveal the signature of magnetic flux emergence/decay in the stellar chromosphere/coronae (e.g. \cite{2019ApJ...871..187N}, \yearcite{2020ApJ...891..103N})

%Future Plan??
\subsection{Future Works}
Finally, we introduce our plan on Seimei-OISTER campaign.
We will continue the flare monitoring on M-dwarf flare stars with Seime-OISTER campaign to obtain statistical samples.
In 2018, Transiting Exoplanet Survey Satellite ($TESS$; \cite{2015JATIS...1a4003R}) was launched and have begun to provide us midterm (27 days - 1 yr) stellar photometric data of nearby stars including flare stars.
This would be a good opportunity to do the long-sought simultaneous observation of stellar flares  as well as rotational modulations  with high-sensitivity space-based photometry gound-based and X-ray telescope, which will open new doors for the stellar flare study.
 Also, we could not detect the line asymmetry of the H$\alpha$ lines, but whether the symmetry/asymmetry is common should be investigated in the future works because it may be related to the stellar coronal mass ejections  (e.g. \cite{2016A&A...590A..11V}, \yearcite{2019A&A...623A..49V}; \cite{2018PASJ...70...62H};  \cite{2018A&A...615A..14F}).  
Moreover, the measurement of the optical spectra for superflares on G-type main-sequence stars (e.g., \cite{2012Natur.485..478M}; \cite{2019ApJ...876..58N}) is still challenging but will be tried by combining $TESS$ with our ground-based observations, which would have an impact on the more understandings on the stellar extreme events.

\bigskip
%Acknowledgement: We acknowledge with thanks H. Hayakawa for their fruitful comments on our work. 
This work was supported by JSPS KAKENHI Grant Numbers
JP16H03955 (K. S.), % (Shibata), 
JP17H02865 (D. N.), % (Nogami),
JP17K05400 (M. H.), % (Maehare),
JP17J08772 (M. K.), %MKimura
JP17K05392 (Y. T.)
and JP18J20048 (K. N.). % (Namekata)
K. N. is supported by the JSPS Overseas Challenge Program for Young Researchers.
Y. N. is supported by the JSPS Overseas Fellowship Program.
 This work was also supported by the Optical and Near-infrared Astronomy Inter-University Cooperation Program and the Grants-in-Aid of the Ministry of Education.

%\begin{figure}[htbp]
%\begin{center}
%\includegraphics[scale=0.5]{SF04_preflare_subtracted_spectra.eps}
%\end{center}
%\caption{}
%\label{fig:lcall}
%\end{figure}

%\begin{figure}[htbp]
%\begin{center}
%\includegraphics[scale=0.7]{SF04_xrayspectra.pdf}
%\end{center}
%\caption{}
%\label{fig:lcall}
%\end{figure}

%\begin{figure}[htbp]
%\begin{center}
%\includegraphics[scale=0.5]{ha_xray_rotation_correlation.eps}
%\end{center}
%\caption{The comparison between H$\alpha$ E.W. and X-ray count rates (0.5-8keV) during non-flaring phase. The squares and crosses show the values where the hardness ratio values (i.e. F(2-8 keV)/F(0.5-2 keV)) are $<$ and $>$ 0.04. }
%\label{fig:rot1}
%\end{figure}

\appendix

\section{Flare Atlas}\label{app:1}

\begin{figure}[htbp]
\begin{center}
\includegraphics[scale=0.4]{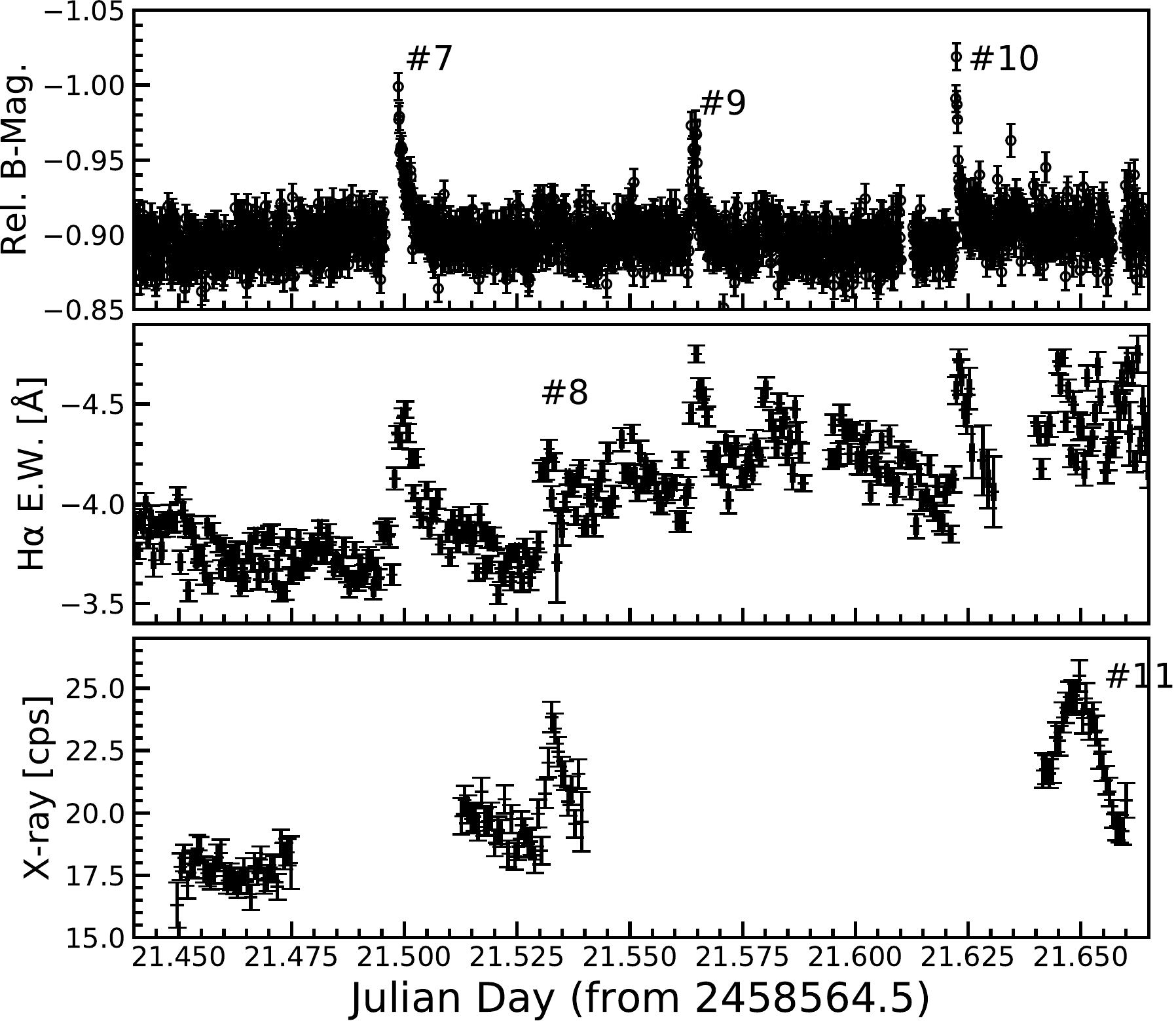}
\end{center}
\caption{B-band, H$\alpha$, X-ray light curve during April 12th. The detected flare \#7 -- \#11 are labeled in the figure.}
\label{fig:app1}
\end{figure}

In this section, we will show the observed flares which are not discussed in the main part but important for our future studies.
Figure \ref{fig:app1} shows the light curves observed with 40cm KU (Kyoto University) Telescope (B-band photometry), Seimei Telescope, and NICER on April 12th.
Five small but clear flares are detected in this period (see flare {\#}7 -- 11).

\begin{figure}[htbp]
\begin{center}
\includegraphics[scale=0.4]{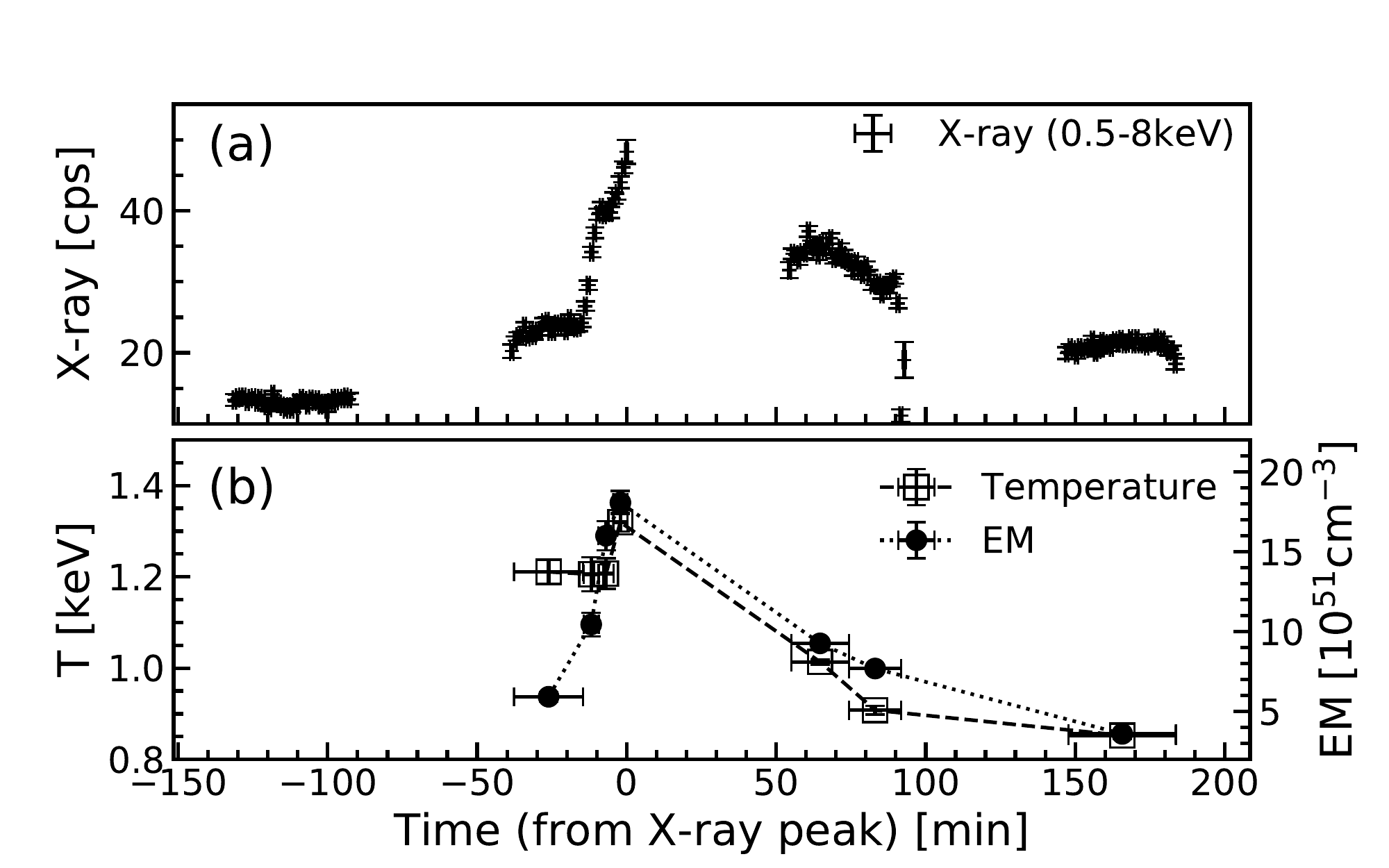}
\end{center}
\caption{X-ray large flares (SF \#12) detected only with NICER. The panel (a) shows the X-ray light curve (0.5 - 8.0 keV). The panel (b) shows the temporal evolution of emission measure and temperature. The radiated energy in 0.5 - 10 keV band is estimated to be 9.7$\times$10$^{31}$ erg. } %9.680118622568599e+31 erg
\label{fig:app2}
\end{figure}

Figure \ref{fig:app2} shows the light curves observed with only NICER on April 13th.
We estimated the emission measure and temperature of the flare \#12, and it is found that the peak timing of temperature is similar to that of the emission measure.
The flare energy in X-ray (0.5 - 10 keV) is estimated to be 9.7 $\times$ 10$^{31}$ erg, which is very large and comparable to the largest scale of solar flares.

\begin{thebibliography}{}
\bibitem[Allred et al.(2005)]{2005ApJ...630..573A} Allred, J.~C., Hawley, S.~L., Abbett, W.~P., et al.\ 2005, \apj, 630, 573

\bibitem[Allred et al.(2006)]{2006ApJ...644..484A} Allred, J.~C., Hawley, S.~L., Abbett, W.~P., et al.\ 2006, \apj, 644, 484

\bibitem[Allred et al.(2015)]{2015ApJ...809..104A} Allred, J.~C., Kowalski, A.~F., \& Carlsson, M.\ 2015, \apj, 809, 104

\bibitem[Allred et al.(2020)]{A20} Allred, J.~C., et al. in preparation

\bibitem[Airapetian et al.(2016)]{2016NatGe...9..452A} Airapetian, V.~S., Glocer, A., Gronoff, G., H{\'e}brard, E., \& Danchi, W.\ 2016, Nature Geoscience, 9, 452 

\bibitem[Aulanier et al.(2013)]{2013A&A...549A..66A} Aulanier, G., D{\'e}moulin, P., Schrijver, C.~J., et al.\ 2013, \aap, 549, A66 

\bibitem[Benz \& G{\"u}del(2010)]{2010ARA&A..48..241B} Benz, A.~O., \& G{\"u}del, M.\ 2010, \araa, 48, 241

\bibitem[Canfield \& Gayley(1987)]{1987ApJ...322..999C} Canfield, R.~C., \& Gayley, K.~G.\ 1987, \apj, 322, 999

\bibitem[Carlsson, \& Stein(1992)]{1992ApJ...397L..59C} Carlsson, M., \& Stein, R.~F.\ 1992, \apjl, 397, L59

\bibitem[Carlsson, \& Stein(1995)]{1995ApJ...440L..29C} Carlsson, M., \& Stein, R.~F.\ 1995, \apjl, 440, L29

\bibitem[Carlsson, \& Stein(1997)]{1997ApJ...481..500C} Carlsson, M., \& Stein, R.~F.\ 1997, \apj, 481, 500

\bibitem[Carlsson, \& Stein(2002)]{2002ApJ...572..626C} Carlsson, M., \& Stein, R.~F.\ 2002, \apj, 572, 626

\bibitem[Crespo-Chac{\'o}n et al.(2006)]{2006A&A...452..987C} Crespo-Chac{\'o}n, I., Montes, D., Garc{\'\i}a-Alvarez, D., et al.\ 2006, \aap, 452, 987

\bibitem[Davenport et al.(2014)]{2014ApJ...797..122D} Davenport, J.~R.~A., Hawley, S.~L., Hebb, L., et al.\ 2014, \apj, 797, 122

\bibitem[Eker(1994)]{1994ApJ...420..373E} Eker, Z.\ 1994, \apj, 420, 373

\bibitem[Fisher(1989)]{1989ApJ...346.1019F} Fisher, G.~H.\ 1989, \apj, 346, 1019

\bibitem[Fuhrmeister et al.(2011)]{2011A&A...534A.133F} Fuhrmeister, B., Lalitha, S., Poppenhaeger, K., et al.\ 2011, \aap, 534, A133

 \bibitem[Fuhrmeister et al.(2018)]{2018A&A...615A..14F} Fuhrmeister, B., Czesla, S., Schmitt, J.~H.~M.~M., et al.\ 2018, \aap, 615, A14 

\bibitem[Gendreau et al.(2016)]{2016SPIE.9905E..1HG} Gendreau, K.~C., Arzoumanian, Z., Adkins, P.~W., et al.\ 2016, \procspie, 99051H

%\bibitem[Graham et al.(2020)]{G20} Graham, D., et al. in revision
  \bibitem[Graham et al.(2020)]{2020arXiv200405075G} Graham, D.~R., Cauzzi, G., Zangrilli, L., et al.\ 2020, arXiv e-prints, arXiv:2004.05075 

\bibitem[Ricker et al.(2015)]{2015JATIS...1a4003R} Ricker, G.~R., Winn, J.~N., Vanderspek, R., et al.\ 2015, Journal of Astronomical Telescopes, Instruments, and Systems, 1, 014003

\bibitem[Rutten(2003)]{2003rtsa.book.....R} Rutten, R.~J.\ 2003, Radiative Transfer in Stellar Atmospheres

\bibitem[Hayakawa et al.(2017)]{2017ApJ...850L..31H} Hayakawa, H., Iwahashi, K., Ebihara, Y., et al.\ 2017, \apjl, 850, L31

\bibitem[Hawley, \& Pettersen(1991)]{1991ApJ...378..725H} Hawley, S.~L., \& Pettersen, B.~R.\ 1991, \apj, 378, 725

\bibitem[Hawley et al.(1995)]{1995ApJ...453..464H} Hawley, S.~L., Fisher, G.~H., Simon, T., et al.\ 1995, \apj, 453, 464

\bibitem[Hawley et al.(2003)]{2003ApJ...597..535H} Hawley, S.~L., Allred, J.~C., Johns-Krull, C.~M., et al.\ 2003, \apj, 597, 535

\bibitem[Hawley et al.(2014)]{2014ApJ...797..121H} Hawley, S.~L., Davenport, J.~R.~A., Kowalski, A.~F., et al.\ 2014, \apj, 797, 121

\bibitem[Heinzel \& Shibata(2018)]{2018ApJ...859..143H} Heinzel, P., \& Shibata, K.\ 2018, \apj, 859, 143

\bibitem[Holman et al.(2003)]{2003ApJ...595L..97H} Holman, G.~D., Sui, L., Schwartz, R.~A., et al.\ 2003, \apjl, 595, L97

\bibitem[Honda et al.(2018)]{2018PASJ...70...62H} Honda, S., Notsu, Y., Namekata, K., et al.\ 2018, \pasj, 70, 62

\bibitem[Hori et al.(1997)]{1997ApJ...489..426H} Hori, K., Yokoyama, T., Kosugi, T., et al.\ 1997, \apj, 489, 426

\bibitem[Hunt-Walker et al.(2012)]{2012PASP..124..545H} Hunt-Walker, N.~M., Hilton, E.~J., Kowalski, A.~F., et al.\ 2012, \pasp, 124, 545

\bibitem[Hussain et al.(2007)]{2007MNRAS.377.1488H} Hussain, G.~A.~J., Jardine, M., Donati, J.-F., et al.\ 2007, \mnras, 377, 1488

\bibitem[Ichimoto \& Kurokawa(1984)]{1984SoPh...93..105I} Ichimoto, K., \& Kurokawa, H.\ 1984, \solphys, 93, 105

\bibitem[Jej{\v{c}}i{\v{c}} et al.(2018)]{2018ApJ...867..134J} Jej{\v{c}}i{\v{c}}, S., Kleint, L., \& Heinzel, P.\ 2018, \apj, 867, 134

\bibitem[Kahler et al.(1982)]{1982ApJ...252..239K} Kahler, S., Golub, L., Harnden, F.~R., et al.\ 1982, \apj, 252, 239

\bibitem[Kotani et al.(2005)]{2005NCimC..28..755K} Kotani, T., Kawai, N., Yanagisawa, K., et al.\ 2005, Nuovo Cimento C Geophysics Space Physics C, 28, 755

\bibitem[Kowalski et al.(2013)]{2013ApJS..207...15K} Kowalski, A.~F., Hawley, S.~L., Wisniewski, J.~P., et al.\ 2013, \apjs, 207, 15

\bibitem[Kowalski et al.(2015)]{2015SoPh..290.3487K} Kowalski, A.~F., Hawley, S.~L., Carlsson, M., et al.\ 2015, \solphys, 290, 3487

\bibitem[Kowalski et al.(2016)]{2016ApJ...820...95K} Kowalski, A.~F., Mathioudakis, M., Hawley, S.~L., et al.\ 2016, \apj, 820, 95

\bibitem[Kowalski et al.(2017)]{2017ApJ...837..125K} Kowalski, A.~F., Allred, J.~C., Uitenbroek, H., et al.\ 2017, \apj, 837, 125

\bibitem[Kowalski et al.(2019)]{2019ApJ...878..135K} Kowalski, A.~F., Butler, E., Daw, A.~N., et al.\ 2019, \apj, 878, 135

\bibitem[Kowalski et al.(2020)]{K20} Kowalski, A.~F., et al. 2020, in preparation

\bibitem[Koyama et al.(1996)]{1996PASJ...48L..87K} Koyama, K., Hamaguchi, K., Ueno, S., et al.\ 1996, \pasj, 48, L87

\bibitem[Krucker et al.(2011)]{2011ApJ...739...96K} Krucker, S., Hudson, H.~S., Jeffrey, N.~L.~S., et al.\ 2011, \apj, 739, 96

 \bibitem[Kurita et al.(2020)]{Kurita2020} Kurita, M., et al.\ 2020, \pasj, accepted 

\bibitem[Lingam \& Loeb(2017)]{2017ApJ...848...41L} Lingam, M., \& Loeb, A.\ 2017, \apj, 848, 41 

\bibitem[Maehara et al.(2012)]{2012Natur.485..478M} Maehara, H., Shibayama, T., Notsu, S., et al.\ 2012, \nat, 485, 478 

\bibitem[Maehara et al.(2017)]{2017PASJ...69...41M} Maehara, H., Notsu, Y., Notsu, S., et al.\ 2017, \pasj, 69, 41 

\bibitem[Marino et al.(2003)]{2003A&A...407L..63M} Marino, A., Micela, G., Peres, G., et al.\ 2003, \aap, 407, L63

\bibitem[Matsubayashi et al.(2019)]{2019PASJ...71..102M} Matsubayashi, K., Ohta, K., Iwamuro, F., et al.\ 2019, \pasj, 71, 102

\bibitem[Matthews et al.(2003)]{2003A&A...409.1107M} Matthews, S.~A., van Driel-Gesztelyi, L., Hudson, H.~S., et al.\ 2003, \aap, 409, 1107

\bibitem[Morin et al.(2008)]{2008MNRAS.390..567M} Morin, J., Donati, J.-F., Petit, P., et al.\ 2008, \mnras, 390, 567

\bibitem[Muheki et al.(2020)]{2020arXiv200306163M} Muheki, P., Guenther, E.~W., Mutabazi, T., et al.\ 2020, arXiv e-prints, arXiv:2003.06163

%\bibitem[Namekata et al.(2017)]{2017ApJ...851...91N} Namekata, K., Sakaue, T., Watanabe, K., et al.\ 2017, \apj, 851, 91.
\bibitem[Namekata et al.(2019)]{2019ApJ...871..187N} Namekata, K., Maehara, H., Notsu, Y., et al.\ 2019, \apj, 871, 187 

\bibitem[Namekata et al.(2020)]{2020ApJ...891..103N} Namekata, K., Davenport, J.~R.~A., Morris, B.~M., et al.\ 2020, \apj, 891, 103

%\bibitem[Namekata et al.(2020)]{2020arXiv200201086N} Namekata, K., Davenport, J.~R.~A., Morris, B.~M., et al.\ 2020, arXiv e-prints, arXiv:2002.01086

\bibitem[Neupert(1968)]{1968ApJ...153L..59N} Neupert, W.~M.\ 1968, \apjl, 153, L59

\bibitem[Notsu et al.(2019)]{2019ApJ...876..58N} Notsu, Y., Maehara, H., Honda, S., et al.\ 2019, \apj, 876, 58

\bibitem[Osten et al.(2006)]{2006ApJ...647.1349O} Osten, R.~A., Hawley, S.~L., Allred, J., et al.\ 2006, \apj, 647, 1349

\bibitem[Paulson et al.(2006)]{2006PASP..118..227P} Paulson, D.~B., Allred, J.~C., Anderson, R.~B., et al.\ 2006, \pasp, 118, 227

\bibitem[Pettersen et al.(1984)]{1984ApJS...54..375P} Pettersen, B.~R., Coleman, L.~A., \& Evans, D.~S.\ 1984, \apjs, 54, 375

\bibitem[Priest(1981)]{1981sfmh.book.....P} Priest, E.~R.\ 1981, Solar Flare Magnetohydrodynamics

\bibitem[Rosner et al.(1978)]{1978ApJ...220..643R} Rosner, R., Tucker, W.~H., \& Vaiana, G.~S.\ 1978, \apj, 220, 643

\bibitem[Saar \& Linsky(1985)]{1985ApJ...299L..47S} Saar, S.~H., \& Linsky, J.~L.\ 1985, \apjl, 299, L47

\bibitem[Schrijver(2020)]{2020ApJ...890..121S} Schrijver, C.~J.\ 2020, \apj, 890, 121

\bibitem[Segura et al.(2010)]{2010AsBio..10..751S} Segura, A., Walkowicz, L.~M., Meadows, V., Kasting, J., \& Hawley, S.\ 2010, Astrobiology, 10, 751 

\bibitem[Shibata, \& Yokoyama(2002)]{2002ApJ...577..422S} Shibata, K., \& Yokoyama, T.\ 2002, \apj, 577, 422

\bibitem[Shibata \& Magara(2011)]{2011LRSP....8....6S} Shibata, K., \& Magara, T.\ 2011, Living Reviews in Solar Physics, 8, 6

\bibitem[Shibata et al.(2013)]{2013PASJ...65...49S} Shibata, K., Isobe, H., Hillier, A., et al.\ 2013, \pasj, 65, 49 

\bibitem[Shkolnik et al.(2009)]{2009ApJ...699..649S} Shkolnik, E., Liu, M.~C., \& Reid, I.~N.\ 2009, \apj, 699, 649

\bibitem[Tei et al.(2018)]{2018PASJ...70..100T} Tei, A., Sakaue, T., Okamoto, T.~J., et al.\ 2018, \pasj, 70, 100

\bibitem[Tremblay \& Bergeron(2009)]{2009ApJ...696.1755T} Tremblay, P.-E., \& Bergeron, P.\ 2009, \apj, 696, 1755

\bibitem[Tsuboi et al.(2016)]{2016PASJ...68...90T} Tsuboi, Y., Yamazaki, K., Sugawara, Y., et al.\ 2016, \pasj, 68, 90

\bibitem[Vida et al.(2016)]{2016A&A...590A..11V} Vida, K., Kriskovics, L., Ol{\'a}h, K., et al.\ 2016, \aap, 590, A11

 \bibitem[Vida et al.(2019)]{2019A&A...623A..49V} Vida, K., Leitzinger, M., Kriskovics, L., et al.\ 2019, \aap, 623, A49 

\bibitem[Wargelin et al.(2017)]{2017MNRAS.464.3281W} Wargelin, B.~J., Saar, S.~H., Pojma{\'n}ski, G., et al.\ 2017, \mnras, 464, 3281

\bibitem[Warren(2006)]{2006ApJ...637..522W} Warren, H.~P.\ 2006, \apj, 637, 522

\bibitem[Watanabe et al.(2017)]{2017ApJ...850..204W} Watanabe, K., Kitagawa, J., \& Masuda, S.\ 2017, \apj, 850, 204


\end{thebibliography}
\end{document}